\documentclass{article}
\usepackage[utf8]{inputenc}
\usepackage{amsmath, amsthm, amssymb, color}
\usepackage[a4paper, total={7in, 10in}]{geometry}
\usepackage[round]{natbib}
\usepackage{float}

\usepackage{prodint}
\usepackage{url}
\usepackage{todonotes} % Paket fuer Notizen
\usepackage[capitalize]{cleveref}

\theoremstyle{definition}
\newtheorem{assumptions}{Assumption}
\newtheorem{theorem}{Theorem}
\newtheorem{remark}{Remark}
\newtheorem{lem}{Lemma}
\newtheorem{ex}{Example}

\renewcommand{\P}{P}
\newcommand{\E}{\mathrm{E}}

\renewcommand{\title}[1]{ \noindent{\centering \Large \textbf{ #1 } \\} }
\newcommand{\inst}[1]{\noindent\textsuperscript{#1}}
\newcommand{\institute}[1]{{\centering \footnotesize{#1}} \vspace{2ex}}

\begin{document}

\title{Effect measures for comparing paired event times}

\begin{center}
        Merle Munko\inst{1,$\ddagger$}\let\thefootnote\relax\footnote{$\ddagger$ Corresponding author. e-mail address: \url{merle.munko@ovgu.de}},
        Simon Mack\inst{2}, 
		Marc Ditzhaus\inst{1,$\dagger$}\let\thefootnote\relax\footnote{$\dagger$ Deceased on September 11, 2024.}, 
        Stefan Fröhling\inst{3,4,5,6}, 
        Dennis Dobler\inst{2,*}\let\thefootnote\relax\footnote{* shared last authorship}, 
        Dominic Edelmann\inst{7,*}

        \vspace{3mm}

        {\today}

\end{center}

	\institute{
        \inst{1} Department of Mathematics, Otto-von-Guericke University Magdeburg, Magdeburg, Germany,\\
        \inst{2} Institute of Statistics, RWTH Aachen University, Aachen, Germany,\\
        \inst{3} National Center for Tumor Diseases (NCT) Heidelberg, Heidelberg, Germany,\\
        \inst{4} Department of Translational Medical Oncology, German Cancer Research Center (DKFZ), Heidelberg, Germany,\\
        \inst{5} German Cancer Consortium (DKTK), Heidelberg, Germany,\\
        \inst{6} Institute of Human Genetics, Heidelberg University, Heidelberg, Germany,\\
        \inst{7} Division of Biostatistics, German Cancer Research Center (DKFZ), Heidelberg, Germany\\
        }

\hrule
\vspace{1cm}
%% Abstract
\noindent\textbf{Abstract:} The progression-free survival ratio (PFSr) is a widely used measure in personalized oncology trials. It evaluates the effectiveness of treatment by comparing two consecutive event times - one under standard therapy and one under an experimental treatment.
However, most proposed tests based on the PFSr cannot control the nominal type I error rate, even under mild assumptions such as random right-censoring. Consequently the results of these tests are often unreliable.
%While these tests naturally work on paired survival data if no censoring is present, they cannot cope with the intricate censoring scheme imposed on the PFSr as a result of right-censored event times.

As a remedy, we propose to estimate the relevant probabilities related to the PFSr by adapting recently developed methodology for the relative treatment effect between paired event times.
As an additional alternative, we develop inference procedures based on differences and ratios of restricted mean survival times. %of paired event times.

An extensive simulation study confirms that the proposed novel methodology  provides reliable inference, whereas previously proposed techniques break down in many realistic settings. The utility of our methods is further illustrated through an analysis of real data from a molecularly aided tumor trial.

\vspace{0.3cm}

%% Keywords

\vspace{1cm}
\hrule
\vspace{0.5cm}

%\Merleinline{Kommentare können mit \textbackslash \textbf{Name}inline\{ Kommentar \} erstellt werden, wobei $\textbf{Name} \in \{\text{Marc, Dennis, Dominic, Simon, Merle}\}$, To-dos mit \textbackslash Todo\{ ... \} }

\section{Introduction}

In medical applications, it is sometimes meaningful to compare two different event times for the same patient. A prominent example are personalized oncology trials  that aim at evaluating the effectiveness of a molecularly guided treatment \citep{massard2017high,sicklick2019molecular,rodon2019genomic,horak2021}. In this context, it has been proposed \citep{von1998there} to compare two consecutive survival times, namely the last time to progression under systemic treatment and progression-free survival under the experimental, molecularly guided treatment.

From a theoretical perspective, such comparisons are naturally framed as paired data problems, since for each individual, we observe two related outcomes $(t_1,t_2)$. Drawing inferences about differences in uncensored paired data is a classical statistical  problem. A well-known approach is the sign test, which investigates if there is the trend for bigger (or smaller) $t_1$ than $t_2$ values. Another prominent example is the exact test by \cite{munzel2002exact} for the null hypothesis of exchangeable pairs; it is based on differences of the ranks within each component.

Whenever the data points represent (positive) event times subject to right-censoring, test statistics for comparing the two marginal samples take less obvious forms.
Most approaches then boil down to comparisons of cumulative hazard or survival functions or constitute variants of log-rank tests; we refer to \cite{woolson1992comparison} for a comparison of different tests and to the introductory section of \cite{dobler24}, which also lists more recent suggestions.

In any case, most approaches developed so far lack the straightforward interpretation of an effect measure.
For this reason, this paper will focus on methods explicitly grounded in interpretable effect measures facilitating the communication with practitioners alongside the mere test decision.

To motivate these effect measures, note that for positive {$(t_1,t_2)$}, their ratio can equivalently be represented as a difference, with the help of the logarithm function: $t_1 / t_2 = \exp( \log(t_1) - \log(t_2))$. 
For this reason, it seems sufficient to focus on ratios of positive event times instead of their differences.
Indeed, several effect measure-based tests \citep{von1998there,mick2000phase,texier2018evaluation,kovalchik2011statistical} for comparing consecutive event times in clinical trials are based on probabilities for such a within-pair ratio,  often called progression-free survival ratio (PFSr) or growth modulation index (GMI).
%As we will see, this concerns the tests proposed by {\color{red}(Mick?! for binomial test? Dominic hatte in Sect. 2 mal etwas von einem Mick geschrieben...)} and \cite{texier2018evaluation,kovalchik2011statistical}.
However, as we will demonstrate, most of these tests are based on rather restrictive assumptions on the type of right-censoring which typically goes hand in hand with a poor performance if these assumptions are not met; see also \cite{edelmann2025progression} and \cite{kuenster2023bachelor}. As a remedy, we will extend recently developed tests by \cite{dobler24} to accommodate the present ratio-focused setting.

As additional alternative, we will develop methodology based on differences and ratios of so-called restricted mean survival times (RMST). These RMSTs have the interpretation of expectations of truncated survival times and recently enjoyed great popularity; see, for example, two-sample tests for RMST differences in \cite{alternative}, permutation tests by \cite{Perm, RMST}, and a factorial designs-based test about contrasts in RMSTs by \cite{munko2024rmst}.

The present paper is organized as follows.
Section~\ref{sec:old_methods} offers a review of various approaches based on or motivated by the PFS ratio.
In Section~\ref{sec:stat_inf}, we will develop our own variant of a PFS ratio-based test by exploiting the tests by \cite{dobler24}, under weaker model assumptions; 
furthermore, we will construct corresponding tests based on RMSTs. 
Our main Theorems~\ref{Normality}--\ref{rmst_randomization} imply that the proposed (asymptotic and randomization/permutation-based) tests are indeed guaranteed to keep the pre-chosen significance level for sufficiently large sample sizes and that these tests are consistent under the alternative hypotheses of interest.
Furthermore, we conducted an extensive simulation study which is described in Section~\ref{sec:simus}.
To make all considered methods comparable, we focused on the effect measure-based confidence intervals and their coverage probabilities.
A real data analysis in Section~\ref{sec:master} concerns a re-analysis of the MASTER trial data about certain patients with advanced cancer and those with rare tumors.
We conclude with a brief discussion in Section~\ref{sec:discuss}.
All proofs are given in the appendix, and an online repository\footnote{https://github.com/dennis-dobler/Effect-measures-for-comparing-paired-event-times} contains RData files with all detailed simulation results.
%\\ \
%{\color{red}[TO DO! Dennis' Github repository.]}

\section{Model and Motivation}
\label{sec:old_methods}

\textcolor{black}{Throughout this paper, we consider independent and identically distributed pairs of survival times $(T_{1i}, T_{2i}), i=1, \dots, n$. 
They model the times to the progression of a disease: first, after the initiation of one treatment, $T_{1i}$ is measured; then, a second treatment phase begins, and $T_{2i}$ is recorded.
In general, $T_{1i}$ and $T_{2i}$ are allowed to be correlated.
Furthermore, we assume that $T_{1i}$ is always observable, whereas the second event time is subject to right-censoring, i.e., we only observe $(T_{1i}, Y_{2i}, \Delta_i), i=1,\dots,n$, where $Y_{2i}=\min(T_{2i}, C_{2i})$ are the censored second event times and $\Delta_i = 1_{\{Y_{2i} \leq C_{2i}\}}$ are the corresponding uncensoring indicators.
Lastly, we assume random right-censoring, i.e., $(T_{1i},T_{2i})$ and $C_{2i}$ are stochastically independent.
At times, we omit the index $i$ when it is not necessary to distinguish the pairs.}

The orginal idea of \cite{von1998there} is to count the treatment of an individual as a ``success'' if the progression-free survival (PFS) ratio $T_2/T_1$ for this individual exceeded a certain threshold $\delta$ for which \cite{von1998there} proposed $1.33$; in later publications most authors chose $\delta = 1.3$, however, other thresholds such as $\delta =1$ or $\delta = 1.5$ were also considered. 
It is now of interest to determine the fraction of ``successes'', e.g., of individuals with $T_2/T_1 \geq \delta$ and corresponding confidence intervals for the probability of this event. % (DD: Dominic, meintest du das?)}.
%\Dennisinline{Soll es tatsächlich $\geq \delta$ sein oder $> \delta$? Vgl.\ andere Literatur.}
% Ist laut Dominic nicht wichtig. Dh der Term mit 1/2 kann auch gerne drin bleiben.
Moreover, one is interested in testing the null hypothesis
\begin{equation} \label{eq:hnullvonhoff}
    H_0 : P(T_2/T_1 \geq \delta) \leq \theta_0,
\end{equation}
where $n \cdot \theta_0$ is the maximum expected number of successes under the null hypothesis (continuation with standard treatment). Alternatively, some authors (e.g. \citealp{texier2018evaluation}) consider testing the hypothesis 
\begin{equation} \label{eq:hnullvonhoff_alt}
    H_0 : P(T_2/T_1 > \delta) \leq \theta_0.
\end{equation}
%Note that it only makes a difference whether~\eqref{eq:hnullvonhoff} or~\eqref{eq:hnullvonhoff_alt} is considered when no ties are pr;
%however, when ties among the event times are present, is does make a difference.
For estimating $P(T_2/T_1 \geq \delta)$ (or $P(T_2/T_1 > \delta)$), various methods have been proposed.

\paragraph{Binomial tests.} The classical approach by \cite{von1998there} estimates  $P(T_2/T_1 \geq \delta)$ via  the number of successes $\sum_{i=1}^n 1_{\{Y_{2i}/T_{1i} \geq \delta\}} = \sum_{i=1}^n 1_{\{\min(T_{2i},C_{2i})/T_{1i} \geq \delta\}}$.  Since this approach treats the censoring time of censored individuals as if it was their survival time, this approach leads to a negatively biased estimator of $P(T_2/T_1 \geq \delta)$. %that is biased downwards. 
Alternatively, it has been proposed to ignore individuals for which the observed censoring time $C_{2i}$  is smaller than  $\delta T_{1i}$ (see e.g., \cite{mick2000phase}).

 In both cases, inference is then carried out using standard binomials tests, which are often markedly conservative; Clopper-Pearson confidence intervals are used.

\paragraph{Kaplan-Meier method.} The idea of the Kaplan-Meier method \citep{kovalchik2011statistical,texier2018evaluation} for estimating $P(T_2/T_1 > \delta)$ is to consider the quantity $T_2/T_1$ as a right-censored survival time and perform estimation of the survival function $S(t) = P(T_2/T_1 > \delta)$ using the Kaplan-Meier estimator $\widehat{S}(\delta)$ for this quantity. In particular, % $\widehat{S}(\delta)$ is used for estimation of  $P(T_2/T_1 > \delta)$
%\Dennisinline{Brauchen wir hier die linksstetige Version von $\hat S$, da $\geq$ und nicht $>$?}
%\Dominicinline{Guter Punkt. Kovalchik beachten diesen Punkt nicht. Texier betrachtet generall $>$. Ich denke, wir sollten an irgendeiner Stelle erwaehnen, dass in der Literatur sowohl $>$ als auch $\geq$ benutzt wird, was bei stetigen Zeiten keinen Unterschied macht, aber in der Realitaet schon. Dann koennen wir auch den $1/2$-Einsatz inhaltlich begruenden}
---corresponding log or complementary log-log-confidence intervals are then used. However, the validity of the Kaplan-Meier estimator crucially depends on the assumption that the censoring times and survival times are independent. For this approach, the (transformed) survival and censoring times are given by $T_2/T_1$ and $C_2/T_1$; due to the common denominator $T_1$, these are virtually never independent. The violation of this assumption canlead to substantial bias of the point estimates and confidence intervals in many settings, see Section~\ref{sec:simus}.

\paragraph{Parametric method based on Weibull frailty model.} \cite{texier2018evaluation} proposed a parametric model for determining point estimates and confidence intervals for $P(T_2/T_1 > \delta)$, assuming that, conditionally on a frailty term $u$, the marginals of $T_1$ and $T_2$ are independently Weibull-distributed with common shape parameter $b$ and scale parameters $u\lambda_1$ and $u\lambda_2$. In particular, conditionally on the frailty term $u$, the following densities for $T_j$, $j \in \{1,2\}$, are assumed
    $$
        f_j (x; b, \lambda_j | u ) = b (u \lambda_j)^{-b} x^{b-1} \exp\left( - \Big(\frac{x}{u \lambda_j} \Big)^b \right), \quad x \geq 0.
    $$
   
The distribution of $T_2/T_1$ then follows a logistic distribution, which is independent of the frailty term,
    $$
        g(x) = b \kappa^{b} x^{b-1}  \left(1- (\kappa x)^b \right)^{-2},
    $$
where $\kappa=\lambda_1/\lambda_2$. Assuming independence of censoring times $C_2/T_1$ and $T_2/T_1$, the maximum likelihood estimators $\widehat{b}$ and $\widehat{\kappa}$ can be easily obtained yielding an estimator for the survival function corresponding to $g(\cdot)$, $S_g(x) = P(T_2/T_1 \geq x)$.

However, as pointed out above, the crucial assumption of independence between $C_2/T_1$ and $T_2/T_1$ will virtually always be violated; additionally, this method makes strong parametric assumptions.
%If the parametric assumptions are not met, it has been shown that the method often performs poorly even in situations where no censoring is present \textcolor{red}{(references)}.

\paragraph{Midrank method.} For explaining the midrank method proposed by \cite{kovalchik2011statistical} and \cite{texier2018evaluation}, let us first assume that $T_2$ and $T_1$ are both uncensored. Define $S_{1i} =\delta T_{1i}$, $S_{2i} =T_{1i}$ and denote the ranks of $S_{ji}$  in the joint sample $(S_{ji})_{j \in \{1,2\}, i \in \{1,\ldots,n\}}$ by  $Q_{ji}$. Then, assuming no ties between $S_{1i}$ and $S_{2i}$, we can estimate $P(T_2/T_1 > \delta)$ via
    $$
        \sum_{i=1}^n 1_{\{Q_{2i} > Q_{1 i} \}} = \sum_{i=1}^n 1_{\{T_{2i}/T_{1i} > \delta\}} = \sum_{i=1}^n 1_{\{S_{2i}/S_{1i} > 1\}},
    $$
which is - in the considered setting without censoring -  equivalent to the classical method described in the paragraph \textit{Binomial tests}.

If $T_2$ is censored, the ranks $Q_{ji}$ are not precisely known for \textit{any} of  the observations (except perhaps for the smallest survival times if they are uncensored); however  one can derive lower bounds $L_{ji}$ and upper bounds $R_{ji}$ for $Q_{ji}$. To obtain a lower bound for the rank  $Q_{kl}$, set $S^{(k,l)}_{kl} = T_{kl}$ and for all  $(i,j) \in (\{1,2\} \times \{1,\ldots,n\}) \backslash \{k,l\}$, set $S^{(k,l)}_{ji} = \Delta_{ji} S_{ji} + (1-\Delta_{ji}) \infty$. Then the lower bound $L_{kl}$ is defined as the rank of $S^{(k,l)}_{kl}$ in the  sample $(S^{(k,l)}_{ji})_{j \in \{1,2\}, i \in \{1,\ldots,n\}}$.
Similarly, to obtain the upper bound $R_{kl}$, we define $U^{(k,l)}_{kl} = \Delta_{kl} S_{kl} + (1-\Delta_{kl}) \infty$ and $U^{(k,l)}_{ji} = S_{ji}$ and define $R_{kl}$ as the rank of $U^{(k,l)}_{kl}$ in the sample $(U^{(k,l)}_{ji})$.

It is now proposed to replace the ranks $Q_{ji}$ by the midranks \citep{hudgens2002midrank},
    $$
        M_{ji} = \frac{L_{ji}+R_{ji}}{2}
    $$
and estimating $P(T_2/T_1 \geq \delta)$ via
$$
   \sum_{i=1}^n 1_{\{M_{2i} \geq M_{1i} \}}. 
$$
Clopper-Pearson confidence intervals are then calculated assuming that the two sequences of midranks are i.i.d. It should be noted that the midrank method as described by \cite{hudgens2002midrank} assumes that the censoring distribution is the same for $S_{1j}$ and $S_{2j}$,
%\Dennisinline{Dominic, was meint "independent of the groups"?}
%\Dominicinline{Denke wir können das einfach so schreiben, wie ich jetzt habe}
an assumption which is violated for the PFS ratio, since there is no censoring for $T_{1i}$. Moreover, the assumption of independence used for the confidence intervals calculation is clearly not satisfied.
%However, the midrank method typically seems to perform better than its competitors described above (references).

\section{Statistical Inference}
\label{sec:stat_inf}

\iffalse
 The data consist of two event times $T_1\sim S_1$ and $T_2\sim S_2$ 
 which are the consecutive times to disease progression under the standard and the experimental treatment, respectively.
 These two times are in general correlated.
 We generally assume that $T_1$ is always observable but that $T_2$ is subject to independent right-censoring.
 Instead, it is also possible to assume independent censoring for $T_1$; see Remark~\ref{indepCensoring} in the appendix for details.
 
 We thus model the observable data as random vectors as follows, where we add the individual-specific subscript $i$ to the previous notation:
 for each individual patient $i\in\{1,\dots, n\}$,
 the observable data point consists of $(T_{1i}, Y_{2i}, \Delta_{i})$ where $Y_{2i} = \min(T_{2i}, C_i)$, $\Delta_{i} = 1\{T_{2i} \leq C_i\}$, and $C_i$ are the right-censoring times that are independent of $(T_1,T_2)$. We assume that $(T_{1i}, Y_{2i}, \Delta_{i}), i\in\{1,...,n\},$ are independent and identically distributed.
 \fi

 %\Dennisinline{Den folgenden Absatz würde ich am Ende eher in die Diskussion stecken.}
 
In this section, we will introduce alternatives to the methods described in Section~\ref{sec:old_methods}.
We remark that no additional assumptions are necessary.
In particular, we allow $S_1$ and $S_2$ to be general survival functions which includes the special cases of continuous and discrete functions. The existence of instantaneous hazard rates, which is commonly assumed in the literature, is not required.

We are going to propose methods that are based on relative and absolute estimands for quantifying the efficacy of an experimental treatment compared to a standard treatment: a variant of the probability as in the von Hoff's method and functions of restricted mean survival times, respectively.

 \subsection{Variant of von Hoff's method}\label{ssec:VonHoff}

% \Todo{weitere Referenzen etc einfuegen}
 
 Von Hoff's method is based on the probability that the ratio of $T_2$ and $T_1$ exceeds a preliminarily chosen threshold~$\delta$.
 The most common choice is $\delta = 1.3$.
 %That is, one is interested in testing the hypotheses:
 %$$ H_0^{\theta}: {\theta} = P(T_2/T_1 > \delta) \leq {\theta}_0 \quad \text{versus} \quad H_1 : {\theta} > {\theta}_0. $$
 Instead of $P(T_2/T_1 > \delta)$ or $P(T_2/T_1 \geq \delta)$, cf.\ \eqref{eq:hnullvonhoff} and \eqref{eq:hnullvonhoff_alt}, we propose to focus on the estimand $$ P(T_2/T_1 > \delta) + \frac{1}{2}P(T_2/T_1 = \delta). $$
The second term, $P(T_2/T_1 = \delta)$, which has the weight $1/2$, is important to take into account that the distribution of $T_2/T_1$ is allowed to have an atom at $\delta$.
Even in the case of continuous $S_1$ and $S_2$, this is possible, as can be seen from the perfectly correlated case $T_2\equiv \delta T_1 $.

 Due to the limited time horizon of studies,  one typically cannot identify this probability. Instead, we consider the estimand
 \begin{align*}
     {\theta} := & P(\min\{T_2,\tau_2\}/\min\{T_1,\tau_1\} > \delta) + \frac{1}{2}P(\min\{T_2,\tau_2\}/\min\{T_1,\tau_1\} = \delta)  \\
     =&P(\min\{T_2,\tau_2\} > \delta \cdot \min\{T_1,\tau_1\}) + \frac{1}{2}P(\min\{T_2,\tau_2\} = \delta \cdot \min\{T_1,\tau_1\}) 
 \end{align*}
 which is closely related to the estimand in  \cite{dobler24}.
 Here, $\tau_1$ and $\tau_2$ denote the maximum follow-up times. %Furthermore, we assume ${\theta}\in (0,1)$.  
 The experimental treatment is then considered effective if that probability exceeds a certain probability ${\theta}_0 \in (0,1)$, the choice of which might depend on the particular medical application.  Thus, we aim to test the hypothesis
 \begin{align}\label{eq:pHypothese}
      H_0^{\theta}: {\theta} \leq {\theta}_0 \quad \text{versus} \quad H_1^{\theta} : {\theta} > {\theta}_0. 
 \end{align}
 In contrast to the original Von Hoff method, we propose an approach that takes the proper handling of right-censoring into account.
 This will lead to an approximately unbiased estimator of ${\theta}$ and, as a consequence, it is expected to improve the reliability and the power of the method.

% To explain why von Hoff's method is biased, let us consider the commonly used statistic  $$X_{n} = \sum_{i=1}^n 1\{ \min\{T_{2i},\tau_2\}/\min\{T_{1i},\tau_1\} > \delta \}\cdot {\Delta_i}  $$ which only involves the uncensored event times. Here, $1\{...\}$ denotes the indicator function of the set $\{...\}$. $X_n$ is used as the test statistic in a binomial test. The binomial test is typically conducted as if $X_n$ is binomially distributed with parameters $n$ (or $n_0 = \sum_{i=1}^n {\Delta_i}$) and ${\theta}$. To revisit these assumptions, let us calculate the expectation, \begin{align*}  E(X_n) & = n \cdot P( \min\{T_{2},\tau_2\}/\min\{T_{1},\tau_1\} > \delta, T_{2} \leq C)  \end{align*} It is clear from this representation that the probability parameter of the binomially distributed random variable $X_n$ fundamentally depends on the distribution of the censoring times.  As a consequence, $X_n/n$ is not an unbiased estimator for ${\theta}$, not even asymptotically as $n\to\infty$. As a consequence, it should not be used to test the hypotheses $H_0^{\theta}$ versus $H_1^{\theta}$.

 Based on the competing risks-based approach in \cite{dobler24}, $\theta$ can be estimated with the help of the Aalen-Johansen estimator \citep{aalen78}.
 To see the connection to the method developed in \cite{dobler24}, 
 define the pair of survival times $(\tilde T_1, \tilde T_2) = (\delta \cdot \min\{T_1,\tau_1\}, \min\{T_2,\tau_2\})$. 
% and the observable data $ (\tilde T_{1i},  Y_{2i},  \Delta_i) $.
 The underlying competing risks data set can be written as
 $$( Z_i, \varepsilon_i ) = (\min\{\delta  T_{1i}, T_{2i}, \tau, C_i\}, \check{\varepsilon}_i  1\{\min\{\delta T_{1i}, T_{2i}, \tau\} \leq C_i\} ), \quad i\in\{1,...,n\},$$
 where $\tau := \min\{\delta \tau_1, \tau_2\}$ and $\check{\varepsilon}_i \in\{1,2,3\}$ denotes the event indicator; see Section~\ref{ssec:proofNormality} in the appendix for details.
 Now, ${\theta}$ can be estimated with the help of the cumulative incidence functions $F_2, F_3$ for type~2 and type~3 events.
 An event of type~$1$ is present if $\tilde T_{1i} > \tilde T_{2i}$ has been \emph{observed}, %, i.e., $\tilde T_{1i} > Y_{2i}$ and $\Delta_i = 1$;
 an event of type~$2$ is present if 
 $\tilde T_{1i} < \tilde T_{2i}$ has been \emph{observed}, %i.e., $\tilde T_{1i} < Y_{i2}$;
 and a type-$3$ event is present if $\tilde T_{1i} = \tilde T_{2i} $ is \emph{observed}.
 In other cases, %where this could not be observed due to censoring, 
 the data point is censored from a competing risks point of view.
Let $\widehat F_2, \widehat F_3$ denote the Aalen-Johansen estimators \citep{aalen78} of a type $2$ and $3$ event, respectively. Since $\theta = F_2(\tau) + \frac{1}{2} F_3(\tau)$, we obtain the estimator 
 \begin{align*}
     \widehat {\theta} = \widehat F_2(\tau) + \frac{1}{2}\widehat F_3(\tau)
 \end{align*} for $\theta$.
 %for ${\theta}$ as in \cite{dobler24}.
 %, where $\tau := \min\{\delta\tau_1,\tau_2\}$. 
Note that one could allow either $\tau_1 = \infty$ or $\tau_2= \infty$ as long as the respectively other terminal time is finite.

 %\begin{assumptions}\label{assumptions} \Merleinline{Ich bin mir noch nicht sicher, welche Annahmen man %braucht (damit $\sigma_p^2, \widetilde\sigma_p^2 >0$).}
 %Throughout this section, we assume
 %    \begin{enumerate}
 %    \item[(1)] ...
        %\item[(1)] $\P ( T_1 > \tau_1, T_2 > \tau_2 ) > 0$, %???
         %\item[(2)] $\P ( T_1 = \tau_1 ) = \P ( T_2 = \tau_2 ) = 0$, %???
         %\item[(3)] $C$ is continuously distributed prior to $\tau$, %???
         %\item[(4)] $\P (C \geq \tau) > 0$ , 
         %\item[(5)] $\P(\min\{\delta T_1, \delta\tau_1\} = \min\{ T_2, \tau_2\}) < 1$
 %    \end{enumerate}
 %\end{assumptions}
To describe the large sample properties of $\widehat \theta$,  let $\xrightarrow{d}$ denote convergence in distribution. 
 An adaptation of Theorems~1 and 2 in \cite{dobler24} justifies the asymptotic normality of the estimation approach under the following assumption.
 \begin{assumptions}\label{assPaired1}
     We assume
     $\P(\delta T_1 \geq \tau, T_2 \geq \tau) > 0$ and $\P(C_2 \geq \tau) > 0$.
 \end{assumptions}
 \begin{theorem}\label{Normality}
    Under Assumption~\ref{assPaired1}, we have $\sqrt{n}(\widehat {\theta} - {\theta})\xrightarrow{d} \mathcal{N}(0,\sigma_\theta^2)$ as $n\to\infty$, where $\sigma_\theta^2$ is defined in Section~\ref{ssec:proofNormality}.
\end{theorem}
 For technical reasons, we need $\sigma_{\theta}^2 > 0$.
 Therefore, we suppose the following.
 \begin{assumptions}\label{assPaired}
 We assume $\sigma_{\theta}^2 > 0$. 
 Under Assumption~\ref{assPaired1}, this is, e.g., the case if at least one of the following holds, which is shown in Lemma~\ref{pdPaired}:
     \begin{enumerate}
         \item[(1)] $\P(T_2 < \min\{\delta T_1, \tau\}) > 0$ and $\P(\tau \leq \min\{\delta T_1,\delta \tau_1\} \leq \min\{T_2,\tau_2\}) > 0$, 
         \item[(2)] $\P(\delta T_1 < \min\{T_2, \tau\}) > 0$ and $\P(\min\{\delta T_1,\delta \tau_1\} \geq \min\{T_2,\tau_2\} \geq \tau) > 0$, 
         \item[(3)] $\P(\delta T_1 = T_2 < \tau) > 0$ and $\P(\min\{\delta T_1,\delta \tau_1\} > \min\{T_2,\tau_2\} > u) \neq \P(u < \min\{\delta T_1,\delta \tau_1\} < \min\{T_2,\tau_2\})$ for all $u\in [0,\tau)$.
     \end{enumerate}
     
 \end{assumptions}
% Note that this assumption also implies $\theta\in (0,1)$.
 
This preliminary work and Slutzky's theorem imply the following result.
\begin{theorem}\label{p_asymptotic}
    Under Assumptions~\ref{assPaired1} and \ref{assPaired}, 
    we have $ \sqrt{n}(\widehat {\theta} - {\theta})/{\widehat \sigma_{\theta}} \xrightarrow{d} \mathcal{N}(0,1)$ as $n\to\infty$, where $\widehat \sigma_{\theta}^2$ is defined as in~\eqref{eq:varestpaired}.
\end{theorem}
 %Theorems~... and ...  in \cite{dobler24} thus justify the asymptotic normality of this estimation approach, as $n\to\infty$, and the validity of the randomization method for inference. 

With this theorem, we can construct an asymptotic level-$\alpha$ test for \eqref{eq:pHypothese}, that is,
\begin{align*}
    \varphi^{\theta} := {1}\left\{ \sqrt{n}(\widehat {\theta} - {\theta}_0)/{\widehat \sigma_{\theta}} > z_{1-\alpha} \right\},
\end{align*} where here and throughout $z_{1-\alpha}$ denotes the $(1-\alpha)$-quantile of the standard normal distribution.
 
% \Merleinline{Ist das der Randomization Approach den wir anschauen wollen?}
Instead of the standard normal quantile, it is typically beneficial to use a resampling-based quantile.
In particular, we propose a randomization approach, i.e.,
%For the randomization approach, 
the observable event indicator $\varepsilon_i$ is randomly re-labeled as 1 or 2 with probability $1/2$, respectively, whenever an event of type 1 or 2 occurred; cf.\ \cite{dobler24} for a similar approach.
This is equivalent to randomly permuting the paired (censored) event times $(\widetilde T_{1i},1)$ and $(Y_{2i},\Delta_i)$ within each pair $i\in\{1,...,n\}$. This results in the randomized data set $(Z_i,\widetilde{\varepsilon}_i), i\in\{1,...,n\},$ and corresponding randomized estimator $\widetilde {\widehat \theta}$. Furthermore, we define the randomization-based variance estimator by $\widetilde \sigma_{\theta}^2$  based on our randomized sample $(Z_i,\widetilde{\varepsilon}_i), i\in\{1,...,n\}.$

Analogously to the proof of Theorem~2 in the supplement of \cite{dobler24}, we obtain that $\sqrt{n}(\widetilde{\widehat\theta} - 1/2) \xrightarrow{d^*} \mathcal{N}(0,\widetilde\sigma_\theta^2)$ conditionally on the data $(Z_i,{\varepsilon}_i), i\in\{1,...,n\},$ in outer probability as $n\to\infty$, where $\widetilde\sigma_\theta^2$ is given in Section~\ref{ssec:proofp_resampling}.
Again, we need to assume a positive variance of the limit.

\begin{assumptions}\label{assPairedresampling}
 We assume $\widetilde\sigma_{\theta}^2 > 0$. 
 Under Assumption~\ref{assPaired1}, this is, e.g., the case if $\P(T_2 < \min\{\delta T_1, \tau\}) >0$ or $\P( \delta T_1 < \min\{T_2, \tau\}) > 0$ holds, which is shown in Lemma~\ref{pdPairedresampling}.     
 \end{assumptions}
Then, we obtain the following result.

\begin{theorem}\label{p_resampling}
    Under Assumptions~\ref{assPaired1} and~\ref{assPairedresampling}, we have $ \sqrt{n}(\widetilde{\widehat\theta} - 1/2)/\widetilde{\widehat \sigma}_{\theta}\xrightarrow{d^*} \mathcal{N}(0,1)$ conditionally on the data $(Z_i,{\varepsilon}_i), i\in\{1,...,n\},$ in outer probability as $n\to\infty$.
    Mathematically, this means
    \begin{align*}
       \sup\limits_{x\in\mathbb R} \left| \P\left( \sqrt{n}(\widetilde{\widehat\theta} - 1/2)/\widetilde{\widehat \sigma}_{\theta} \leq x \mid (Z_i,{\varepsilon}_i), i\in\{1,...,n\} \right) - \Phi(x) \right| \xrightarrow{P} 0
    \end{align*} as $n\to\infty$, where $\Phi: \mathbb R \to [0,1]$ denotes the cumulative distribution function of the standard normal distribution.
\end{theorem}
%The proof of this theorem follows analogously to the proof of Theorem~2 in the supplement of \cite{dobler24}.
%
Theorem~\ref{p_resampling} provides that the randomization test 
\begin{align*}
    \widetilde\varphi^{\theta} := {1}\left\{ \sqrt{n}(\widehat {\theta} - {\theta}_0)/{\widehat \sigma_{\theta}} > \widetilde z_{1-\alpha} \right\}
\end{align*} is an asymptotic level $\alpha$ test, where $\widetilde z_{1-\alpha}$ denotes the $(1-\alpha)$-quantile of the conditional distribution of $\sqrt{n}(\widetilde{\widehat\theta} - 1/2)/{\widetilde{\widehat\sigma}_{\theta}}$ given the data $(Z_i,{\varepsilon}_i), i\in\{1,...,n\}.$ In practice, the quantile $\widetilde z_{1-\alpha}$  can be approximated by a Monte Carlo method. 

 By means of standard arguments, the classical bootstrap, i.e., randomly drawing with replacement from the three-dimensional data points $(T_{1i},Y_{2i},\Delta_i),$ $i\in\{1,...,n\}$, can also be shown to be valid, e.g., by applying Theorem~3.7.1 in \cite{vaartWellner2023}. Indeed, it also exhibited a good behaviour in the small sample simulation study in \cite{dobler24}, albeit not quite as good as the randomized approach.
 To make the bootstrap approach more explicit, we denote the bootstrap versions of the estimators with an additional superscript ${}^*$.
Thus, a bootstrap version of the test statistic above is $\sqrt{n} (\widehat \theta^* - \widehat \theta)/ \widehat{\sigma}^*_\theta$.
 %\Merleinline{Soll hier noch mehr zum Bootstrap hin?}
 %\Dennisinline{Hängt vielleicht von der Zielgruppe/Journal ab. Und von ersten Simulationsergebnissen... (Ist der Bootstrap viel besser? Dann eher ja. Im SiM-Papier von Kathrin und mir war Randomisation eher besser.)}
 
 \subsection{Restricted mean survival times} \label{sec:rmst}
 %\Marcinline{Hier ergänze ich später mal ein paar Zitate für den RMST, u.a. Merles Papier.}
 An alternative approach for comparing paired survival times is the comparison of the restricted mean survival times (RMSTs) of the two event times. The RMST is defined as the area under the survival curve up to a prespecified time point $\tau > 0$, that is,
$$ \mu_j := \int_0^{\tau} S_j(t) \;\mathrm{d}t \in [0,\tau], \quad j\in\{1,2\}. $$
For comparing two RMSTs, we can consider the hypotheses
$$ H_0^{\text{diff}}: \mu_2 - \mu_1 \leq \eta \quad \text{versus} \quad H_1^{\text{diff}}: \mu_2 - \mu_1 > \eta $$
for the difference of the RMSTs with $\eta \in [-\tau,\tau]$, or
$$ H_0^{\text{rat}}: \frac{\mu_2}{\mu_1} \leq 1 + \zeta \quad \text{versus} \quad H_1^{\text{rat}}: \frac{\mu_2}{\mu_1} > 1 + \zeta$$
for the ratio of RMSTs with $\zeta \in (-1,\infty)$. A natural estimator for $\mu_j$ is given by
$$ \widehat \mu_j := \int_0^{\tau} \widehat S_j(t) \;\mathrm{d}t $$
for $j\in\{1,2\}$, where $\widehat S_j$ denotes the Kaplan-Meier estimator of $S_j$.
Hence, we get the estimator $\widehat \mu_2 - \widehat \mu_1$ for $\mu_2 - \mu_1$ and $\widehat \mu_2 / \widehat \mu_1$ for $\mu_2/ \mu_1$.

 For technical reasons, we need the following assumptions.
% assume ${\sigma}_{\text{diff}}, {\sigma}_{\text{rat}} > 0$, where ${\sigma}_{\text{diff}}, {\sigma}_{\text{rat}}$ are defined as in Section~\ref{ssec:proof3} in the appendix.
 \begin{assumptions}\label{ass_rmst} %\Merleinline{Ich bin mir noch nicht sicher, welche Annahmen man braucht, damit die Varianz $>$ 0.}
 %\Dennisinline{Erstmal Ass1 ohne die sigmas, dann Theorem mit ggf. Konvergenz gegen 0, schliesslich Ass2 mit sigmas.} \Merleinline{erledigt ;)}
 Throughout this section, we assume
     \begin{enumerate}
     %\item[(1)] ${\sigma}_{\text{diff}}^2, {\sigma}_{\text{rat}}^2 > 0$, where ${\sigma}_{\text{diff}}^2, {\sigma}_{\text{rat}}^2$ are defined as in Section~\ref{ssec:proof3} 
     %\item[(1)] $\mu_1,\mu_2 > 0$, %damit man durch mu teilen darf
     \item[(1)] $\P (C_2 \geq \tau) > 0$ and%damit man durch H_j teilen darf
     \item[(2)] $\P (T_1 \geq \tau),\P (T_2 \geq \tau) > 0$. %damit man durch S_j teilen darf
%     \item[(3)] either $\min\{T_1,\tau\}$ or $\min\{T_2,\tau\}$ is not deterministic, %damit man Varianz in den Daten hat und was vor tau beobachtet
%     \item[(4)] $\P(\min\{T_{1},\tau\} \neq \min\{X_2,\tau\}) > 0$ or $\P(C < \tau) > 0$. %??? %P(C<tau) ist equivalent zu P(C < min{T_2,tau}) = P(Zensierung vor tau) unter (2) und unabhängigkeit
     %damit die Schätzer für die Gruppen nicht immer gleich sind = Varianz null haben
     %\item[(5)] $\P( \mathrm{IF}^{\text{diff}}_1 \neq 0 ) > 0$ or $\P( \mathrm{IF}^{\text{rat}}_1 \neq 0 ) > 0$, respectively.
     \end{enumerate}
 \end{assumptions}

Under the stated assumptions, the estimators can be shown to be asymptotically normal, that is
\begin{align*}
    &\sqrt{n}((\widehat \mu_2 - \widehat \mu_1) - ( \mu_2 - \mu_1)) \xrightarrow{d} \mathcal{N}(0,{\sigma}_{\text{diff}}^2) \\
    \text{and }\quad &\sqrt{n}(\log(\widehat \mu_2 / \widehat \mu_1) - \log(\mu_2 /\mu_1))\xrightarrow{d} \mathcal{N}(0,{\sigma}_{\text{rat}}^2)
\end{align*} as $n\to\infty$ for some ${\sigma}_{\text{diff}}^2,{\sigma}_{\text{rat}}^2 \geq 0$; see Section~\ref{ssec:proof3} in the appendix for details. %In the following, we further assume ${\sigma}_{\text{diff}}^2, {\sigma}_{\text{rat}}^2 > 0$.
 
\begin{theorem}\label{rmst_asymptotic}
    Under Assumptions~\ref{ass_rmst} and ${\sigma}_{\text{diff}}^2, {\sigma}_{\text{rat}}^2 > 0$, we have $$ \sqrt{n}((\widehat \mu_2 - \widehat \mu_1) - ( \mu_2 - \mu_1))/\widehat{\sigma}_{\text{diff}} \xrightarrow{d} \mathcal{N}(0,1)$$
    and
    $$\sqrt{n}(\log(\widehat \mu_2 / \widehat \mu_1) - \log(\mu_2 /\mu_1))/\widehat{\sigma}_{\text{rat}}\xrightarrow{d} \mathcal{N}(0,1)$$ as $n\to\infty$. The definitions of the variance estimators $\widehat{\sigma}^2_{\text{diff}}$ and $\widehat{\sigma}^2_{\text{rat}}$ are given in Section~\ref{ssec:proof3} in the appendix.
\end{theorem}

This theorem yields that the tests
\begin{align*}
    \varphi^{\text{diff}} &:= 1\left\{  \sqrt{n}((\widehat \mu_2 - \widehat \mu_1) - \eta )/\widehat{\sigma}_{\text{diff}} > z_{1-\alpha} \right\}\\
    \text{and }\quad \varphi^{\text{rat}} &:= 1\left\{\sqrt{n}(\log(\widehat \mu_2 / \widehat \mu_1) - \log(1+\zeta))/\widehat{\sigma}_{\text{rat}}  > z_{1-\alpha}\right\}
\end{align*}
are asymptotic level $\alpha$ tests for $ H_0^{\text{diff}}$ and $ H_0^{\text{rat}}$, respectively.

The randomization approach described in Section~\ref{ssec:VonHoff} can be adopted to construct a randomization test. 
To this end, let $(Y_{1i}^{\pi}, \Delta_{1i}^{\pi})$ and $(Y_{2i}^{\pi},  \Delta_{2i}^{\pi})$ denote the permuted (censored) event times of the paired (censored) event times $(T_{1i}, 1)$ and $(Y_{2i},\Delta_i)$ within each pair $i\in\{1,...,n\}.$
Furthermore, we denote all estimators based on the permuted (censored) event times $(Y_{1i}^{\pi}, \Delta_{1i}^{\pi}),(Y_{2i}^{\pi},  \Delta_{2i}^{\pi}), i\in\{1,...,n\},$ with a $\pi$ in the superscript in the following. E.g., 
$\widehat \mu_j^{\pi}$  denotes the RMST estimator based on the permuted (censored) event times  $(Y_{ji}^\pi, \Delta_{ji}^\pi), i\in\{1,...,n\},$ for $j\in\{1,2\}$.
The following theorem yields the consistency of this randomization approach.

\begin{theorem}\label{rmst_randomization}
    Under Assumption~\ref{ass_rmst} and ${\sigma}_{\text{diff}}^{\pi},{\sigma}_{\text{rat}}^{\pi}>0$, we have, as $n\to\infty$,
    $$ \sqrt{n}(\widehat \mu_2^{\pi} - \widehat \mu_1^{\pi}) /\widehat{\sigma}_{\text{diff}}^{\pi} \xrightarrow{d^*} \mathcal{N}(0,1)$$
    and
    $$\sqrt{n}\log(\widehat \mu_2^{\pi} / \widehat \mu_1^{\pi})/\widehat{\sigma}_{\text{rat}}^{\pi}\xrightarrow{d^*} \mathcal{N}(0,1)$$  conditionally on the data $(T_{1i},Y_{2i},\Delta_{2i}), i\in\{1,...,n\}$ in outer probability, where
    ${\sigma}_{\text{diff}}^{\pi},{\sigma}_{\text{rat}}^{\pi}$ are defined in Section~\ref{ssec:proof3}.
\end{theorem}

Hence, the validity of the randomization tests, that are,
\begin{align*}
    \varphi^{\pi,\text{diff}} &:= {1}\left\{  \sqrt{n}((\widehat \mu_2 - \widehat \mu_1) - \xi )/\widehat{\sigma}_{\text{diff}} >  z_{1-\alpha}^{\pi,\text{diff}} \right\}\\
    \text{and }\quad \varphi^{\pi,\text{rat}} &:= {1}\left\{\sqrt{n}(\log(\widehat \mu_2 / \widehat \mu_1) - \log(1+\zeta))/\widehat{\sigma}_{\text{rat}}  >  z_{1-\alpha}^{\pi,\text{rat}}\right\},
\end{align*} 
is provided, where $z_{\alpha}^{\pi,\text{diff}}$ and $z_{\alpha}^{\pi,\text{rat}}$ denote the $(1-\alpha)$-quantiles of the conditional distributions of $\sqrt{n}(\widehat \mu_2^{\pi} - \widehat \mu_1^{\pi}) /\widehat{\sigma}_{\text{diff}}^{\pi}$ and $\sqrt{n}\log(\widehat \mu_2^{\pi} / \widehat \mu_1^{\pi})/\widehat{\sigma}_{\text{rat}}^{\pi}$, respectively, given the data $(T_{1i}, Y_{2i}, \Delta_i), i\in\{1,...,n\}.$
The quantiles can also be approximated by a Monte Carlo method.

Finally, we would like to remark  again that an application of the bootstrap would similarly lead to an asymptotically level-$\alpha$ test.
The bootstrap versions of the test statistics are 
$\sqrt{n} ((\widehat \mu_2^* - \widehat \mu_1^*) - (\widehat \mu_2 - \widehat \mu_1))/ \widehat \sigma_{\text{diff}}^*$ and $\sqrt{n} (\log(\widehat \mu_2^*/\widehat \mu_1^*) - \log(\widehat \mu_2/\widehat \mu_1))/ \widehat \sigma_{\text{rat}}^*$, respectively.

%\Merleinline{Soll hier auch noch was zum classical Bootstrap hin?}
%\Dennisinline{Warten wir erstmal die Gespräche mit den anderen ab. :-)}

\section{Simulation study} \label{sec:simus}
\subsection{Methods}
We compare our proposed methods to the Kaplan-Meier and Midrank method. Von Hoff's approach is not included in the simulation study, as its poor performance, which is in line with the theoretical shortcomings, was already demonstrated in a simulation study by \citet{edelmann2025progression}. We also omitted the Weibull-Frailty model, due to the strong parametric model assumptions.
\subsection{Data Generation}
The survival times are generated using the bivariate Weibull distribution described by \citet{lu1990some}, with joint survival function
\begin{equation}
    S(t_1,t_2)=\exp{\left(-\left((t_1/\theta_1)^{\frac{\gamma_1}{\nu}}+(t_2/\theta_2)^{\frac{\gamma_2}{\nu}}\right)^{\nu}\right)}
\end{equation}
where $\theta_1, \theta_2$ and $\gamma_1,\gamma_2$ are positive real valued parameters which correspond to scale and shape parameters of the marginal Weibull distributions respectively. The parameter $\nu \in (0,1]$ models the dependence, where $\nu=1$ equals independence and the correlation increases with decreasing $\nu$. \citet{lu1990some} also give an explicit formula for the joint moments of $(T_1,T_2)$, from which the correlation can be obtained in closed form. As the resulting expression is highly complicated, especially for the case of unequal shape parameters $\gamma_1,\gamma_2$, we omit it here. We chose $\delta=1$ and $\delta=1.3$ for the RTE, Kaplan-Meier and Midrank methods. Additionally, we used the cutoff times $\tau_1=1.2$ and $\tau_2=\delta\tau_1.$ The RMST-based methods are only included in the first setting, i.e., $\delta=1$, as they are designed for equal cutoff times in both groups.

\begin{figure}[H]
    \centering
	\includegraphics[width=0.7\textwidth]{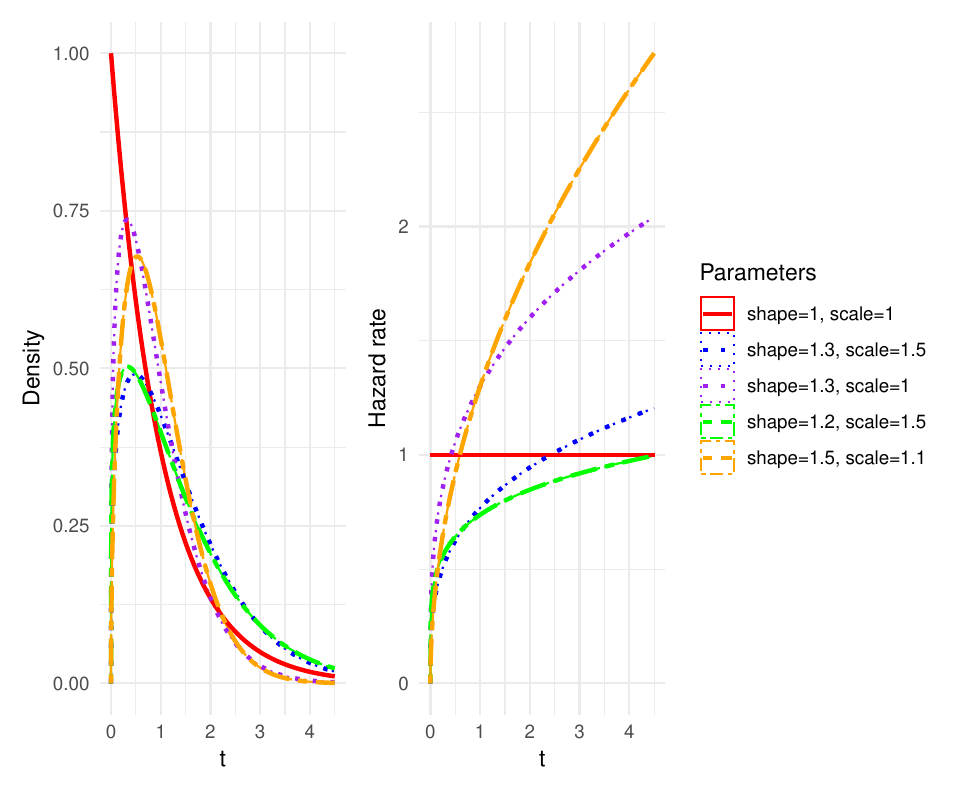}
	\caption{Density and hazard functions of considered Weibull marginal distributions}
\label{fig:densities}
\end{figure}

Regarding the marginal distributions of $T_1$ and $T_2$, we considered three different sets of parameters. First, we consider equal marginal distributions, with the parameters $\theta_1=\theta_2=\gamma_1=\gamma_2=1.$ In this case, $T_1$ and $T_2$ are standard exponentially distributed. For the second setting, we chose $\theta_1=1.5, \theta_2=1$ and $\gamma_1=\gamma_2=1.3,$ which implies a constant hazard ratio between the two survival times of 
\[\text{HR}:=\left(\frac{\theta_1}{\theta_2}\right)^{\gamma_1}=1.5^{1.3}\approx1.694.\]
We also consider the case, that both scale and shape of the marginal distributions are different, by choosing the parameters $\theta_1=1.5, \theta_2=1.1$ and $\gamma_1=1.2, \gamma_2=1.5.$ In this setting the hazard ratio is time varying. Figure \ref{fig:densities} illustrates the density as well as hazard functions of the considered marginal distributions.

All scenarios were simulated for sample sizes $n\in\{20,30,50,100,200\}$ and values of the dependence parameter $\nu\in \{1,0.8,0.6,0.4\}.$ Censoring times were generated from exponential distributions with rate parameters $\lambda=1/2$ (medium censoring) and $\lambda=1$ (heavy censoring). Uncensored event times were considered as well. All simulations were conducted in the R computing environment, version 4.4.0 \citep{citeR} each with $n_{\text{sim}}=5000$ simulation runs. The random quantiles in the permutation procedures were determined via $B=1000$ random permutations.

\subsection{Results}
The methods were compared in terms of coverage probabilities of left-sided 95\% confidence intervals, as the corresponding tests do not necessarily test the same hypothesis. Coverage probabilities for right- and two-sided intervals were obtained as well, and are in most scenarios very similar. The detailed simulation results for each setting are available in RData format in an online repository.\footnote{https://github.com/dennis-dobler/Effect-measures-for-comparing-paired-event-times}
For the ratio of $\text{RMSTs}$, the left-sided asymptotic $(1-\alpha)$-confidence interval is given by
\[\left[\frac{\widehat \mu_2} {\widehat \mu_1}\exp{(-n^{-1/2}\widehat{\sigma}_{\text{rat}}z_{1-\alpha})}, \infty\right),\]
the two-sided and permutation intervals are defined analogously.
The true values of $\theta$ and $P(T_2>T_1)$ were obtained as an estimate from a sample of 100,000 uncensored observations, as their exact computation would involve integrating over the joint distribution of $(T_1,T_2).$ The true values of $\text{RMST}_1$ and $\text{RMST}_2$ have been obtained trough numerical integration.

\begin{figure}[h!]
    \centering
	\includegraphics[width=0.82\textwidth]{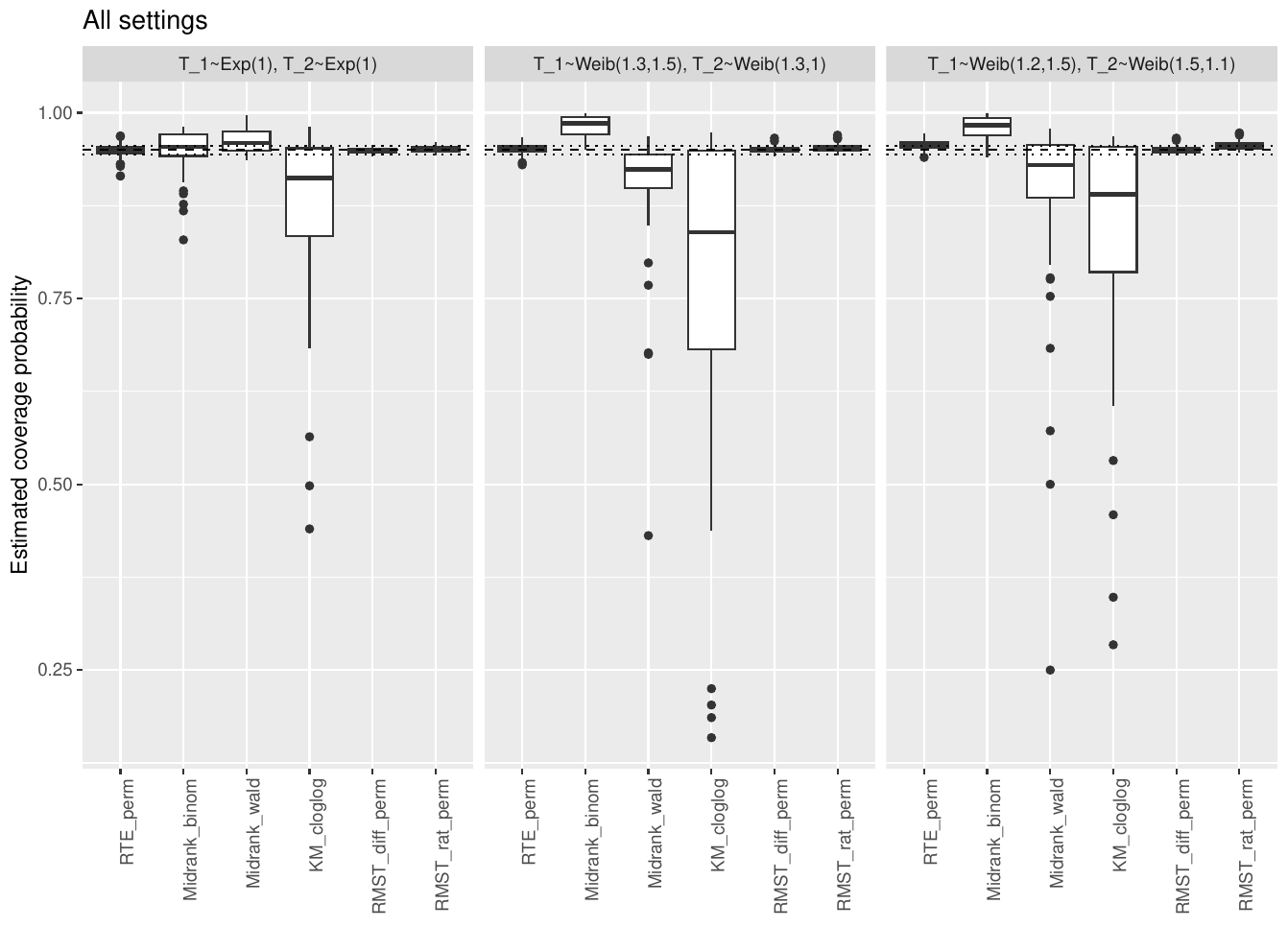}
	\caption{Estimated coverage probabilities over all settings of left-sided 95\% confidence intervals for $\delta=1$, stratified by marginal distribution. The dashed lines represent the borders of the binomial confidence interval $[94.4\%, 95.6\%]$.}
\label{fig:res_full_left}
\end{figure}

\begin{figure}[h!]
    \centering
	\includegraphics[width=0.82\textwidth]{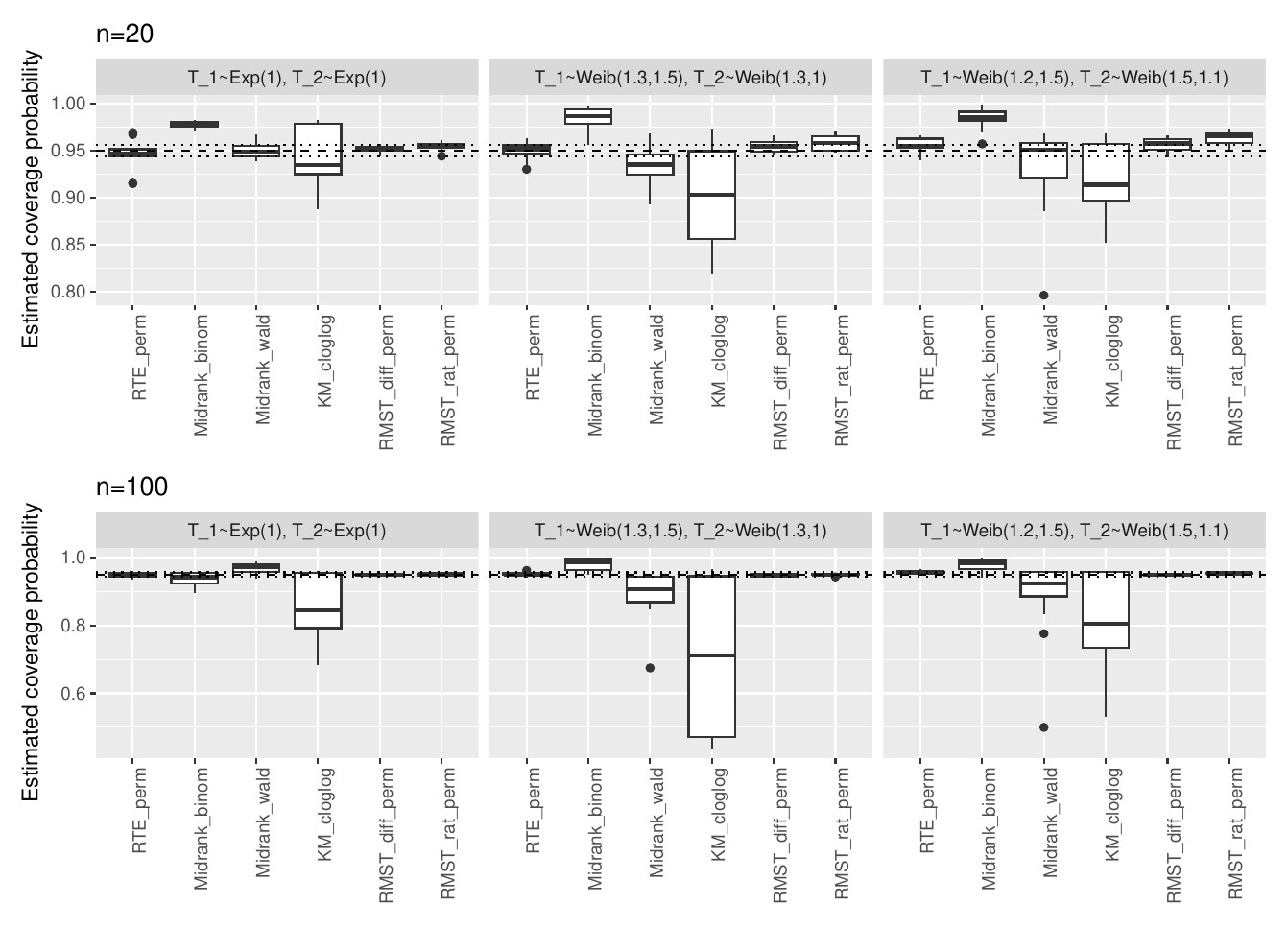}
	\caption{Estimated coverage probabilities for different sample sizes of left-sided 95\% confidence intervals for $\delta=1$, stratified by marginal distribution. The dashed lines represent the borders of the binomial confidence interval $[94.4\%, 95.6\%]$.}
\label{fig:res_samplesize_left}
\end{figure}

\begin{figure}[h!]
    \centering
	\includegraphics[width=0.82\textwidth]{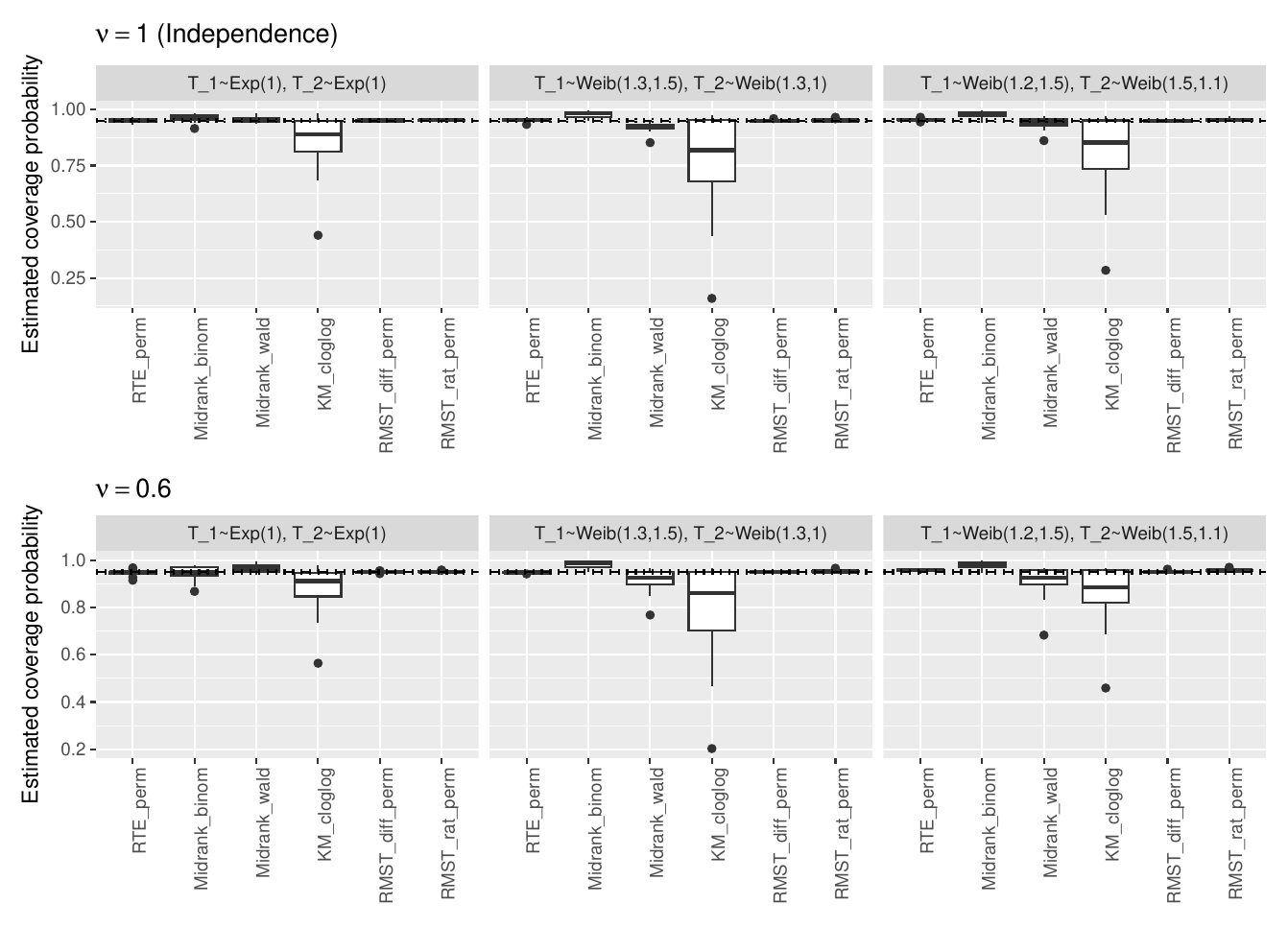}
	\caption{Estimated coverage probabilities for different dependence parameters of left-sided 95\% confidence intervals for $\delta=1$, stratified by marginal distribution. The dashed lines represent the borders of the binomial confidence interval $[94.4\%, 95.6\%]$.}
\label{fig:res_dependence_left}
\end{figure}

\begin{figure}[h!]
    \centering
	\includegraphics[width=0.82\textwidth]{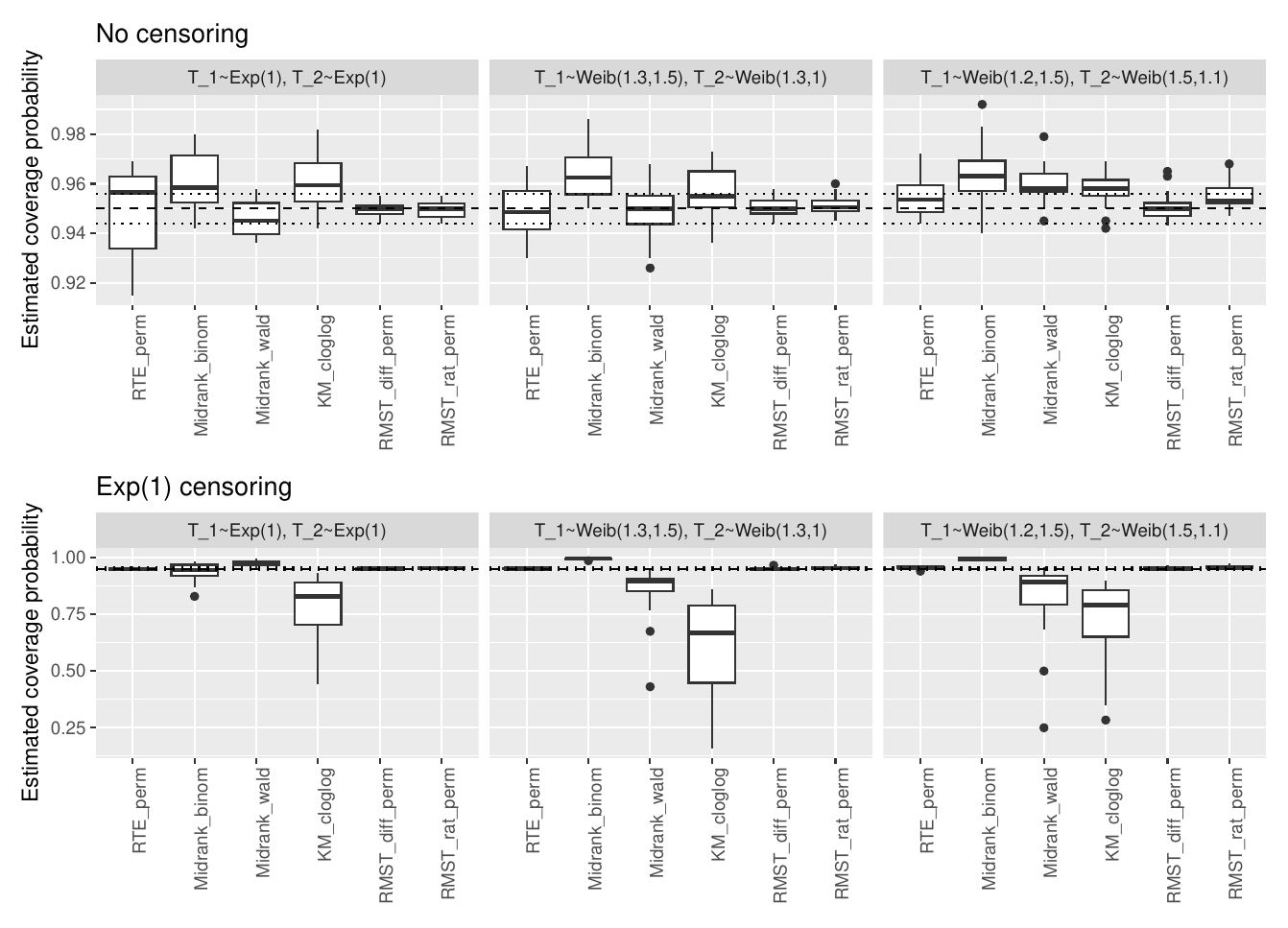}
	\caption{Estimated coverage probabilities for different censoring distributions of left-sided 95\% confidence intervals for $\delta=1$, stratified by marginal distribution. The dashed lines represent the borders of the binomial confidence interval $[94.4\%, 95.6\%]$.}
\label{fig:res_cens_left}
\end{figure}

Figure \ref{fig:res_full_left} illustrates the overall results for $\delta=1$.  Figures \ref{fig:res_samplesize_left}, \ref{fig:res_dependence_left}, and \ref{fig:res_cens_left} display subsets of the overall results for different simulation parameters, again for the scenario $\delta=1.$ The nominal coverage is controlled well by our proposed methods, with a small but noticeable conservative behavior of the RTE and the ratio of $\text{RMSTs}$ in the case of non-proportional hazards. Only the permutation versions of the intervals are presented, as the performance is similar to the asymptotic intervals in larger samples, and superior for small to medium sample sizes. The Kaplan-Meier method fails to control the nominal level in most scenarios, displaying substantial under-coverage and by far the largest variability between settings. This can be attributed to the dependence between $T_2/T_1$ and $C_2/T_1$, as already noted in Section \ref{sec:old_methods}. Only if no censoring is present, the coverage is close to the nominal level. For the Midrank method, the results are twofold. The binomial (Clopper-Pearson) intervals are in most cases overly conservative, whereas the Wald intervals are too liberal. The different directions of the bias may be attributed to the fact that Wald intervals are based on a studentization approach, whereas binomial intervals are not.

Even in the case of small sample sizes of $n=20,$ or high censoring ($\text{Exp}(1)$ distributed censoring times lead to censoring rates of over 50\%) our proposed methods produce coverage probabilities close to the nominal level, even if the marginal distributions are different. The midrank method seems to struggle especially with unequal marginals, whereas the coverage is close to the nominal level if the marginals are equal. Figures \ref{fig:res_n20_left}-\ref{fig:res_exp1_left} in appendix \ref{sec:appendix_simus} confirm these findings. For the Kaplan-Meier and Midrank methods, the coverage probabilities depart further from the nominal level with increasing dependence and censoring, and partly also with increasing sample size, exemplifying their inconsistency.

\begin{figure}[h!]
    \centering
	\includegraphics[width=0.82\textwidth]{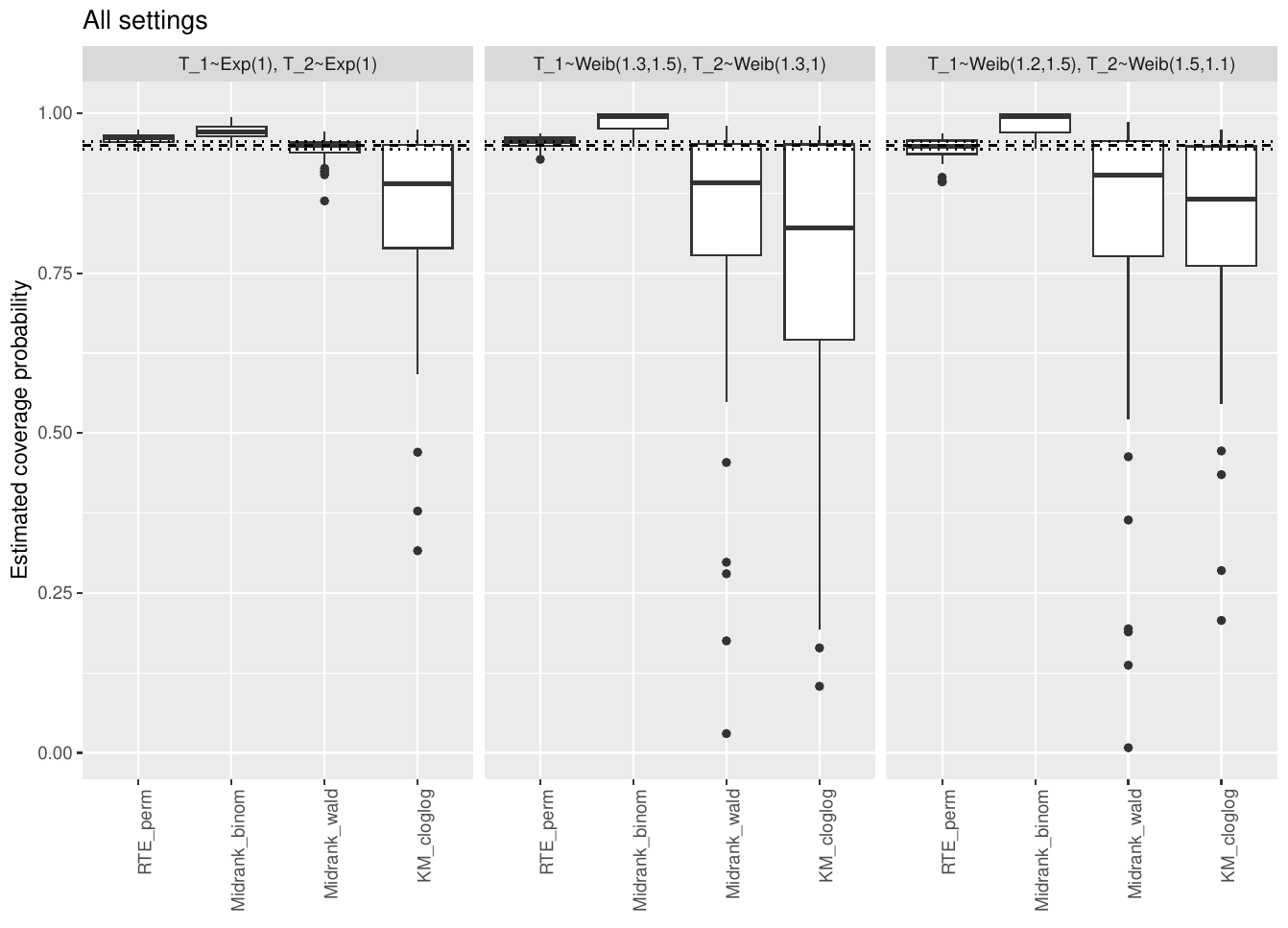}
	\caption{Estimated coverage probabilities over all settings of left-sided 95\% confidence intervals for $\delta=1.3$, stratified by marginal distribution. The dashed lines represent the borders of the binomial confidence interval $[94.4\%, 95.6\%]$.}
\label{fig:gesamt_left_delta1.3}
\end{figure}

In the scenario with $\delta=1.3$ the results are very similar, we therefore only illustrate the overall results in Figure \ref{fig:gesamt_left_delta1.3}. The RTE again controls the nominal level well, with a slightly conservative behavior for equal marginal distributions. The Midrank method combined with the Wald interval also works well for equal marginal distributions; however for unequal marginals the method is again too liberal, whereas the binomial intervals are too conservative. The Kaplan-Meier method again fails to control the nominal level for all marginal distributions. These findings are in line with the results obtained for $\delta=1.$ However, a noticeable difference to the previous scenario is present for the, not displayed, right- and two-sided intervals, here the RTE is more liberal in some scenarios and larger samples are needed to reach the nominal level, a behavior not noticeable for $\delta=1.$ The RTE is preferable in this case nonetheless, as the behavior of the Midrank and Kaplan-Meier method is similar to the displayed results for left-sided intervals.

\newpage
\phantom{.}
\newpage

\section{Real data analysis: re-analysis of the MASTER trial data}
\label{sec:master}

We now apply the proposed methods and competing approaches to a real data example from the MASTER (Molecularly Aided Stratification for Tumor Eradication Research) trial \citep{horak2017, horak2021}. MASTER is a multicenter, prospective observational study, which includes young adults ($< 51$ years) with advanced cancer and patients with rare tumors, regardless of age. Patients in MASTER undergo a standardized workflow, involving molecular profiling and bioinformatic analyses.  Results are then discussed in a multidisciplinary molecular tumor board, where molecularly informed recommendations  are given. These recommendations may lead to diagnostic reevaluation, genetic counseling or experimental therapy. 

Our analyses focus on a subset of $n=238$ patients who received molecularly informed treatment and had undergone at least one prior systemic therapy, for which the time to progression under the last prior systemic therapy was available; cf.\ \cite{edelmann2025progression}. Clinical and demographical information for these patients are provided in Table~\ref{tab:tab1}. 

\begin{figure}[h!]
    \centering
	\includegraphics[width=0.6\textwidth]{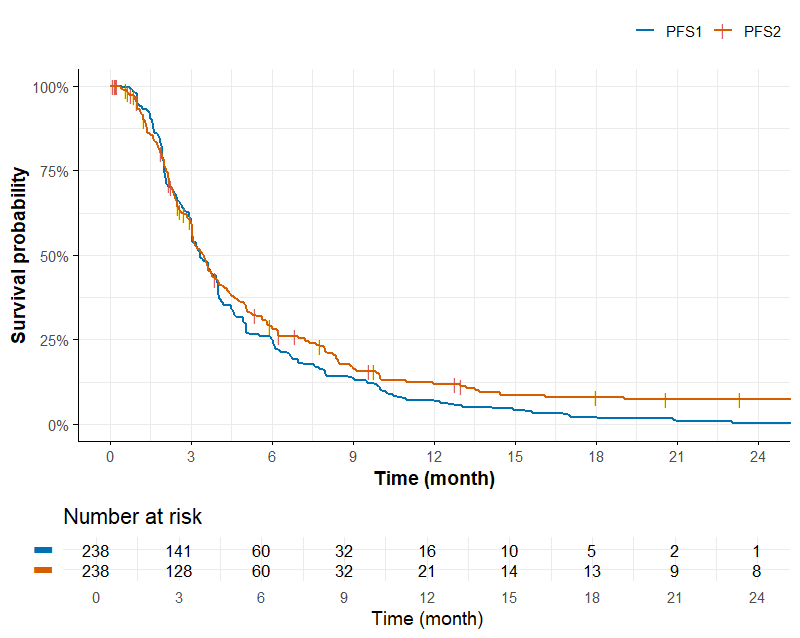}
	\caption{Kaplan-Meier plots for the time to progression under the last prior systemic therapy ($T_1$, in blue) and progression-free-survival under the molecularly informed treatment ($T_2$, in red). Censoring times are marked with a vertical bar.}
\label{fig:km}
\end{figure}

We are interested in comparing the time to progression under the last prior systemic therapy ($T_1$) with progression-free-survival under the molecularly informed treatment ($T_2$). The distributions of $T_1$ and $T_2$ are illustrated via Kaplan-Meier plots in Figure~\ref{fig:km}.

We first apply our method for estimating 
    $$
 P(T_2/T_1 > \delta) + \frac{1}{2}P(T_2/T_1 = \delta),
    $$  
 presented in Section~\ref{ssec:VonHoff} with $\delta \in \{1,1.3\}$ and compare the results with the method of  \cite{von1998there}, the midrank approach \citep{kovalchik2011statistical,texier2018evaluation}, and the Kaplan-Meier method \citep{kovalchik2011statistical,texier2018evaluation}. Different confidence intervals were calculated for the methods and resampling-based confidence intervals were derived from $B=$10,000 resampled data sets. We set $\tau_1 = 1 \, \text{year}$ and $\tau_2 = \delta \, \tau_1$.  The results for $\delta = 1$ and $\delta = 1.3$ are given in Tables~\ref{tab:delta1} and~\ref{tab:delta13}, respectively.

Von Hoff’s approach yields a substantially smaller estimate than the other approaches, a behavior that can likely be attributed to its built-in conservativeness, as discussed in detail in \cite{edelmann2025progression}. Our methods, the Kaplan-Meier method, and the midrank approach give very similar results. However, the Kaplan-Meier and the midrank approach are theoretically flawed and may show very unsatisfactory performance in terms of CI coverage whereas our methods feature asymptotic guarantees and perform substantially better in simulations (see Section~\ref{sec:simus}). We also note a near perfect agreement of the permutation-based and asymptotic CIs, providing additional justification for the use of the asymptotic CIs for applications.

We next apply the methods derived in Section~\ref{sec:rmst} for comparing the RMSTs $\mu_1$ of $T_1$ and $\mu_2$ of $T_2$. Point estimates for the RMST difference $\mu_2 - \mu_1$ and RMST ratio $\mu_2/\mu_1$ with corresponding $95 \%$ CIs are provided in Table~\ref{tab:rmst}; permutation-based CIs were obtained from $B=$ 10,000 permutations. Similar to the relative treatment effect, the asymptotic CIs show excellent agreement with the permutation-based counterparts.

\begin{table}[h!]
\footnotesize
\centering
\caption{Clinical and demographical information on the MASTER program patients considered in our real-world data example.}
\begin{tabular}{ll|rrr}
 \textbf{Variable} & \textbf{Levels} & $\mathbf{n}$ & $\mathbf{\%}$ & $\mathbf{\sum \%}$ \\ 
  \hline
Sex & female & 110 & 46.2 & 46.2 \\ 
   & male & 128 & 53.8 & 100.0 \\ 
   \hline
 & all & 238 & 100.0 &  \\ 
   \hline
Age in years &  17-29 & 30 & 12.6 & 12.6 \\ 
 & 30-49 & 124 & 52.1 & 64.7\\ 
& $ \geq 50$ & 84 & 35.3 & 100.0 \\   
   \hline
 & all & 238 & 100.0 &  \\ 
\hline
Best response & Complete Response & 5 & 2.1 & 2.1 \\ 
   & Mixed response & 14 & 5.9 & 8.0 \\ 
   & Partial response & 39 & 16.4 & 24.4 \\ 
   & Progressive disease & 73 & 30.7 & 55.0 \\ 
   & Stable disease & 45 & 18.9 & 73.9 \\ 
   & Unknown & 62 & 26.1 & 100.0 \\ 
   \hline
 & all & 238 & 100.0 &  \\ 
   \hline

\label{tab:tab1}
\end{tabular}
\end{table}

\begin{table}[h!]
\footnotesize
\centering
\begin{tabular}{lrrrrr}
\hline
Method (CI type) & Estimate & 95\% CI (two-sided)  & 95\% CI (one-sided)    \\
\hline
RTE (Gauss)         & 0.479 & [0.413, 0.544] & [0.424, 1] \\
RTE  (perm)          & 0.479 & [0.413, 0.544] & [0.424, 1] \\
Midrank (binomial)      & 0.471 & [0.406, 0.536] & [0.416, 1]\\
Midrank (Wald)      & 0.471 & [0.407, 0.534] &[0.417, 1] \\
KM (loglog) & 0.478 & [0.411, 0.542] & [0.422, 1]\\
Von Hoff (binomial)     & 0.437 & [0.373, 0.503] & [0.383, 1] \\
\hline
\end{tabular}
\caption{Point estimate and $95 \%$ CIs for relative treatment effect estimates for the real data setting of the MASTER trial with $\delta =1$.}
\label{tab:delta1}
\end{table}

%\vspace{15mm}

\begin{table}[h!]
\footnotesize
\centering
\begin{tabular}{lrrrrr}
\hline
Method (CI type) & Estimate & 95\% CI (two-sided)  & 95\% CI (one-sided)    \\
\hline
RTE (Gauss)         & 0.389 & [0.325, 0.453] & [0.335, 1] \\
RTE  (perm)          & 0.389 & [0.324, 0.453] & [0.335, 1] \\
Midrank (binomial)      & 0.382 & [0.320, 0.447] & [0.330, 1]\\
Midrank (Wald)      & 0.382 & [0.320, 0.444] & [0.331, 1] \\
KM (loglog) & 0.388 & [0.324, 0.452] & [0.334, 1] \\

Von Hoff (binomial)     & 0.340 & [0.280, 0.404] & [0.289, 1]\\
\hline
\end{tabular}
\caption{Point estimate and $95 \%$ CIs for relative treatment effect estimates for the real data setting of the MASTER trial with $\delta =1.3$.}
\label{tab:delta13}
\end{table}

\begin{table}[h!]
\footnotesize
\centering
\begin{tabular}{lrrrrr}
\hline
Method (CI type) & Estimate & 95\% CI (two-sided)  & 95\% CI (one-sided)    \\
\hline
RMST-Diff (Gauss)         & $10.79$ & $[-6.91, 28.49]$ & $[-4.06, \infty)$ \\
RMST-Diff   (perm)          & $10.79$ & $[-7.05, 29.19]$ & $[-3.91, \infty)$ \\
RMST-Ratio  (Gauss)      & $1.080$ & $$[0.953, 1.224]$$ & $[0.972, \infty]$\\
RMST-Ratio  (perm)      & $1.080$ & $[0.952, 1.229]$ & $[0.973, \infty)$ \\
\hline
\end{tabular}
\caption{Point estimate and $95 \%$ CIs for RMST-based procedures for the real data setting of the MASTER trial.}
\label{tab:rmst}
\end{table}

\newpage
\phantom{.}
\newpage
%\phantom{.}
%\newpage

\section{Discussion}
\label{sec:discuss}

This paper demonstrated that many statistical procedures with a focus on progression free survival ratios are flawed and we provided explanations for this.
On the other hand, we verified that our novel alternatives based on the relative treatment effect and restricted mean survival times are reliable remedies.
Mathematical proofs of their large sample properties and simulation results under various settings underlined their usefulness.
While we did not observe great differences between the existing methods and the probability-based novel method in the conducted real data analysis, it is still important to know for the applying statistician that a method has certain reliability guarantees.

It is important to note that this study is merely a first step toward more sophisticated statistical techniques.
One limitation of the present methodology is that it is fully nonparametric, and it would not easily scale with a large number of additional covariates.
As a solution, combinations of the present techniques and partly parametric models could be investigated.
For example, such approaches could be based on proportional subdistribution hazard models \citep{fine99} for the cumulative incidence functions defining the relative treatment effect, cause-specific proportional hazards models \citep{cox72}, or other regression approaches involving inverse-probability-of-censoring weighting \citep{scheike08} or the pseudo-observation approach \citep{andersen03}.

\section*{Acknowledgments}
Merle Munko and Marc Ditzhaus gratefully acknowledge support from the Deutsche Forschungsgemeinschaft (Grant No. DI 2906/1-2 and GRK 2297 MathCoRe). Dennis Dobler and Simon Mack wish to thank their former affiliations TU Dortmund University and Research Center Trustworthy Data Science and Security (University Alliance Ruhr) as well as Vrije Universiteit Amsterdam (Dobler) and Otto-von-Guericke University Magdeburg (Mack) where a smaller part of the work was done.
 
 \bibliographystyle{plainnat}
 \bibliography{main.bbl}

\appendix
\section{Proofs}

\begin{remark}\label{indepCensoring}
    Instead of assuming that the first survival time $T_1$ is always uncensored, we also may assume independent censoring, i.e., $(C_1,C_2)$ is stochastically independent of $(T_1,T_2)$. 
    %\Dennisinline{Ist dieser Fall realistisch?}
    %\Merleinline{Ich denke hier z.B. an das Beispiel mit der Erblindung vom linken und rechten Auge - könnte das nicht so ein Fall sein?}
    This case is even more general since we can set $C_{1} = \tau_1 + 1$ for the case where $T_1$ is uncensored. Hence, all proofs were conducted for the independent censoring case in the following. 
    Therefore, we set $\Delta_j := 1\{T_j \leq C_j\}$ and $Y_j := \min\{T_j,C_j\}$. As before, an additional subscript $i\in\{1,...,n\}$ represents the data of individual $i$.
    
    However, one should be aware that, for some data sets, it might not make sense to assume $(C_1, C_2)$ to be independent of $(T_1,T_2)$. So, the choice to use this modeling approach should be carefully reasoned. 
    % Das folgende habe ich jetzt mal auskommentiert, weil es mich nicht mehr so richtig überzeugt hat bzw. weil das unseren Ansatz eigentlich auch kaputt machen würde.
    \iffalse
Note that for studies in which $T_1$ is also subject to right-censoring which affects both event times, 
 the censoring cannot be assumed independent of the event times:
 since there is only right-censoring time, say $\tilde C$, 
 $T_1$ is compared with $\tilde C$ but $T_2$ is compared with $\tilde C - T_1$, if $T_1 < \tilde C$ and $T_2$ is the time between the first and second event.
 We thus see that the right-censoring time $C= \tilde C - T_1$ for $T_2$ is generally not independent through the correlation between $T_1$ and $T_2$.
\fi

\end{remark}

\subsection{Proof of Theorem~\ref{Normality}}\label{ssec:proofNormality}
As in \cite{dobler24}, let 
\begin{align*}
    \check{\varepsilon}  := \begin{cases}
        1 & \text{if } \min\{\delta T_1, \delta\tau_1\} >  \min\{T_2, \tau_2\} \\
        2 & \text{if } \min\{\delta T_1, \delta\tau_1\} <  \min\{T_2, \tau_2\} \\
        3 & \text{if } \min\{\delta T_1, \delta\tau_1\} =  \min\{T_2, \tau_2\} 
    \end{cases}
\end{align*} denote the (uncensored) event indicator, $\check{T} := \min\{\delta T_1, T_2, \tau\}$ and $\check{C} := \min\{\delta C_1, C_2\}$. Thus, we can write the censored competing risks data set as
$$( Z_i, \varepsilon_i ) = (\min\{\check{T}_i,\check{C}_i\}, \check{\varepsilon}_i  1\{\check{T}_i \leq \check{C}_i\} ), \quad i\in\{1,...,n\}.$$
Furthermore, let
$F_j(t) := \P (\check{T} \leq t, \check{\varepsilon} = j), 
    S(t)   := \P (\check{T} > t)$, $A_j(t) := \int_{[0,t]} \frac{1}{S_-(u)} \;\mathrm{d}F_j(u)$, $A(t) := A_1(t) + A_2(t) + A_3(t)$ and $G(t) := \P (\check{C} > t)$
for all $t \geq 0, j \in\{1,2,3\}$, where here and throughout $M_-$ denotes the left-continuous version of a monotone function $M$.
Note that
\begin{align*}
    F_2(\tau) &= \P (\delta \min\{T_1, \tau_1\} < \min\{T_2, \tau_2\}) \\
    \text{and }\quad F_3(\tau) &= \P (\delta \min\{T_1, \tau_1\} = \min\{T_2, \tau_2\}).
\end{align*}

Firstly, we emphasize that the Aalen--Johansen estimator $\widehat{A}_j$ consistently estimates the cause-specific cumulative hazard function ${A}_j$, i.e., no relevant information is lost
by the above-described competing risks data. Regarding Theorem~4.2 in \cite{dobler2017}, we need to show the following statement.

\begin{lem}\label{Am}
    It holds $A_j(t) = \int_{[0,t]} \frac{1}{\P(Z_1 \geq .)} \;\mathrm{d}\P(Z_1 \leq ., \varepsilon_1 = j)$ for all $j\in\{1,2,3\}, t\geq 0$ with $G_-(t) > 0.$
\end{lem}
\begin{proof}[Proof of Lemma~\ref{Am}]
    Let $j\in\{1,2,3\}, t\geq 0$ be arbitrary with $G_-(t) > 0.$
    Due to the definition of $Z_1$, we have 
    \begin{align*}
        \P(Z_1 \geq u) = \P(\check{T}_1\geq u, \check{C}_1\geq u) = \P(\check{T}_1\geq u)\P(\check{C}_1\geq u) = S_-(t)G_-(t)
    \end{align*} for all $u\in [0,t]$.
    Moreover, 
    \begin{align*}
        \P(Z_1 \leq u, \varepsilon_1 = j) = \P(\check{T}_1 \leq u, \check{\varepsilon}_1 = j, \check{C}_1 \geq \check{T}_1) = \int_{[0,u]} G_-\;\mathrm{d}F_j
    \end{align*} for all $u\in [0,t]$.
    Hence, it follows
    \begin{align*}
        \int_{[0,t]} \frac{1}{\P(Z_1 \geq .)} \;\mathrm{d}\P(Z_1 \leq ., \varepsilon_1 = j)
        = \int_{[0,t]} \frac{G_-}{S_-G_-} \;\mathrm{d}F_j = \int_{[0,t]} \frac{1}{S_-} \;\mathrm{d}F_j  = A_j(t).
    \end{align*}
\end{proof}

To analyze the asymptotic behavior of $\widehat\theta$, we do not use the results of \cite{dobler24} for two reasons: First, we suppose weaker assumptions on the survival and censoring distributions and, second, the variance formula given in \cite{dobler24} contains a flaw which is corrected in the recent version of Dennis Dobler's GitHub repository;\footnote{\url{https://github.com/dennis-dobler/relative_treatment_effect_paired_survival/blob/main/variance_calculations_for_Dobler_and_Moellenhoff_2024_SiM_v2.pdf}}
more details can be found in Section~3.3 in \cite{Share_it_1981185920-123226}.
The mentioned flaw is also illustrated in the following examples. 
%are both wrongly stated, as the following examples show. 
%In the GitHub repository, there is just a bracket missing, such that the second and third stated summand of the final result should be multiplied with $1/4$ as well. Then, the formula is correct under $\Delta F_1(\tau) = \Delta F_2(\tau) = 0$, which follows from the assumptions in \cite{dobler24}. 
However, since we also aim to allow mass in $\tau$ for all event types, we state a different variance formula later. Our formula coincides with the formula in the GitHub repository under $\Delta F_1(\tau) = \Delta F_2(\tau) = 0$.

\begin{ex}
    Let $$ F_1(t) := 0, \quad F_2(t) := \begin{cases}
        0 & \text{if $t < 1$}\\ 1/2 & \text{if $t \geq 1$}
    \end{cases}, \quad\text{and } F_3(t) := \begin{cases}
        0 & \text{if $t < 2$}\\ 1/2 & \text{if $t \geq 2$}
    \end{cases} $$ for all $t\geq 0$ with $\tau = 2$ and $C_1,C_2 \geq 2$ almost surely. Then,
    one can show $\widehat\theta =  \frac{1}{n}\sum_{i=1}^n\left({1}\{\check{\varepsilon}_i = 2\}+  \frac{1}{2}{1}\{\check{\varepsilon}_i = 3\}\right)$ and, thus, by the central limit theorem, $\sqrt{n}(\widehat{\theta}-\theta) \xrightarrow{d} \mathcal{N}(0,1/16)$.
    However, the formula in the supplement of \cite{dobler24} yields 
    \begin{align*}
        \sigma_\theta^2 &= %S_-(1)S_-(1)\sigma_2^2(1)\Delta A_2(1)\Delta A_2(1) + S_-(1)S_-(2)\sigma_2^2(1)\Delta A_2(1)\Delta A_3(2) \\&\quad + S_-(2)S_-(2)\left(\frac{\Delta\sigma_2^2(1)}{1 - \Delta A_2(1)}\left(\frac{1}{2}\Delta A_3(2)\right)^2 - 2\frac{\Delta\sigma_2^2(1)}{1 - \Delta A_2(1)}\left(\frac{1}{2}\Delta A_3(2)\right)\Delta A_3(2) + \frac{1}{4}\sigma_3^2(2)\Delta A_3(2)\Delta A_3(2) \right)
        %\\&= 
        \frac{1}{4}\cdot \frac{1}{2}\cdot\frac{1}{2} + \frac{1}{2}\cdot\frac{1}{4}\cdot\frac{1}{2}\cdot\frac{1}{2} + \frac{1}{2}\cdot\frac{1}{2}\cdot\left(\frac{1/4}{1 - 1/2}\cdot\left(\frac{1}{2}\right)^2 - 2\frac{1/4}{1 - 1/2}\cdot\frac{1}{2} + 0 \right) = 0
    \end{align*}
    if the integrals $\int_0^\tau$ are meant as $\int_{[0,\tau]}$.     \hfill \qed
    \iffalse
    The formula of the GitHub repository yields
    \begin{align*}
        \sigma_\theta^2 &= %\frac{(F_2(1) - F_{2,-}(2) - F_1(1) + F_{1,-}(2) + S(1))^2}{(1 - \Delta A_2(1))^2}\Delta\sigma_2^2(1)
        %= 
        \frac{(1/2 - 1/2  + 1/2)^2}{(1 - 1/2)^2}\cdot \frac{1}{4} =  \frac{1}{4}.
    \end{align*}
    Note that this is not correct due to the missing factor $1/4$.
    \fi
\end{ex}
Also if the integrals $\int_0^\tau$ are meant as $\int_{[0,\tau)}$, the formula in the supplement of \cite{dobler24} is not correct, as the following example shows.
\begin{ex}
    Let $$ F_1(t) := \begin{cases}
        0 & \text{if $t < 1$}\\ 1/2 & \text{if $t \geq 1$}
    \end{cases}, \quad F_2(t) := 0, \quad\text{and } F_3(t) := \begin{cases}
        0 & \text{if $t < 2$}\\ 1/2 & \text{if $t \geq 2$}
    \end{cases} $$ for all $t\geq 0$ with $\tau = 2$ and $C_1,C_2 \geq 2$ almost surely. Then,
    one can show that $\widehat\theta = \frac{1}{2} \frac{1}{n}\sum_{i=1}^n {1}\{\check{\varepsilon}_i = 3\}$ and, thus, by the central limit theorem, $\sqrt{n}(\widehat{\theta}-\theta) \xrightarrow{d} \mathcal{N}(0,1/16)$.
    However, the formula in the supplement of \cite{dobler24} yields $\sigma_\theta^2 = 0$ if the integrals $\int_0^\tau$ are meant as $\int_{[0,\tau)}$.
    \hfill \qed
\end{ex}

Now, to prove Theorem~\ref{Normality},
    note that $\theta$ and $\widehat{\theta}$ can be written as 
\begin{align*}
    \tilde{\psi}\left(\tilde{\phi}\left(-(A_1+A_2+A_3)\right)_-, A_2 + \frac{1}{2}A_3\right)(\tau) &= \int_{[0,\tau]} S_- \;\mathrm{d}\left(A_2 + \frac{1}{2}A_3\right) \\&= F_2(\tau) + \frac{1}{2}F_3(\tau) = \theta
\end{align*}
and
\begin{align*}
    \tilde{\psi}\left(\tilde{\phi}\left(-(\widehat{A}_1+\widehat{A}_2+\widehat{A}_3)\right)_-, \widehat{A}_2 + \frac{1}{2}\widehat{A}_3\right)(\tau) &= \int_{[0,\tau]} \widehat{S}_- \;\mathrm{d}\left(\widehat{A}_2 + \frac{1}{2}\widehat{A}_3\right) \\&= \widehat{F}_2(\tau) + \frac{1}{2}\widehat{F}_3(\tau) = \widehat{\theta}
\end{align*}
with $\tilde{\psi}:\tilde{D}[0,\tau]\times BV_M[0,\tau] \to D[0,\tau], \tilde{\phi}:BV_{3M}[0,\tau] \to {D}[0,\tau)$ as in \cite{DeltaMathod} for some $M <\infty$.
Here, $BV_M[0,\tau]$ denotes the set of all real-valued càdlàg functions on $[0,\tau]$ with total variation bounded by $M$,
$D[0,\tau], D[0,\tau)$ denote the sets of all real-valued càdlàg functions on $[0,\tau], [0,\tau)$, respectively, and
$\tilde{D}[0,\tau]$ denotes the set of all real-valued càglàd functions on $[0,\tau]$.
Then, we define
\begin{align*}
    &\Psi: (BV_M[0,\tau])^3 \to \mathbb R, \\& \Psi(\Lambda_1,\Lambda_2,\Lambda_3) := \tilde{\psi}\left(\tilde{\phi}\left(-(\Lambda_1+\Lambda_2+\Lambda_3)\right)_-, \Lambda_2 + \frac{1}{2}\Lambda_3\right)(\tau).
\end{align*}
To apply the delta-method, we show the Hadamard differentiability of $\Psi$ at $(A_1,A_2,A_3)$ by the chain rule.
Note that Assumption~\ref{assPaired1} is equivalent to $S_-(\tau) > 0$ and $\P(Z_1 \geq \tau) > 0$.
Analogously to Lemma~3.10.18 and Lemma~3.10.32 in \cite{vaartWellner2023}, we obtain the Hadamard-derivatives
\begin{align*}
    &\tilde{\psi}^{\prime}_{(\tilde{\phi}(-(A_1+A_2+A_3))_-,A_{2}+\frac{1}{2}A_3)}(\alpha,\beta) %= \int_{[0,.]} \phi(-A_i) \;\mathrm{d}\beta + \int_{[0,.]} \alpha \;\mathrm{d}A_{im} 
    = \int_{[0,.]}  \tilde{\phi}(-(A_1+A_2+A_3))_-\;\mathrm{d}\beta + \int_{[0,.]} \alpha \;\mathrm{d}\left(A_{2}+\frac{1}{2}A_3\right)
\end{align*} and \begin{align*}
    &\tilde{\phi}^{\prime}_{-(A_1+A_2+A_3)}\left( \beta \right) %= \int_{[0,.)} \prodi_{0\leq x < u}(1 - \mathrm{d} A_i(x)) \prodi_{u < x < .}(1 - \mathrm{d} A_i(x)) \;\mathrm{d}\beta(u)
    = \tilde{\phi}(-(A_1+A_2+A_3))(.) \int_{[0,.]} \frac{1}{1 - \Delta(A_1+A_2+A_3)} \;\mathrm{d}\beta
\end{align*} for all $\alpha\in \Tilde{D}[0,\tau],\beta\in D[0,\tau]$ under Assumption~\ref{assPaired1}. 
Here, we consider $\tilde{\phi}:BV_M[0,\tau) \to D[0,\tau)$ as function mapping to  $D[0,\tau)$ instead of $D[0,\tau]$ to guarantee that the weaker assumption $S_-(\tau)>0$ instead of $S(\tau)>0$ suffices, cf. \cite{DeltaMathod}.
Moreover, $(BV_M[0,\tau])^3\ni(\Lambda_1,\Lambda_2,\Lambda_3) \mapsto \Lambda_1 + \Lambda_2 + \Lambda_3$, $(BV_M[0,\tau])^3\ni(\Lambda_1,\Lambda_2,\Lambda_3) \mapsto \Lambda_2 + \frac{1}{2}\Lambda_3$, $D[0,\tau) \ni \Lambda \mapsto \Lambda_- \in \tilde D[0,\tau]$ and $D[0,\tau] \ni \Lambda \mapsto \Lambda(\tau)\in\mathbb R$ are linear and, thus, their Hadamard-derivatives equals the functionals, respectively.
Hence, the chain rule implies that $\Psi$ is Hadamard differentiable at $(A_1,A_2,A_3)$ with Hadamard-derivative 
\begin{align*}
    &\Psi'_{(A_1,A_2,A_3)}(\alpha_1,\alpha_2,\alpha_3) \\=& \tilde{\psi}'_{(\tilde{\phi}\left(-(A_1+A_2+A_3)\right)_-, A_2 + \frac{1}{2}A_3)}\left( \tilde{\phi}'_{-(A_1+A_2+A_3)}\left(-(\alpha_1 + \alpha_2 + \alpha_3)\right)_- , \alpha_2 + \frac{1}{2}\alpha_3\right)(\tau)
    \\=& \int_{[0,\tau]} \tilde{\phi}\left(- (A_1,A_2,A_3)\right)_- \;\mathrm{d}\left(\alpha_2 + \frac{1}{2}\alpha_3\right) \\ &- \int_{[0,\tau)} \frac{\int_{(u,\tau]}\tilde{\phi}\left(- (A_1,A_2,A_3)\right)_-\;\mathrm{d}\left(A_2 + \frac{1}{2}A_3\right)}{1-\Delta (A_1+A_2+A_3)(u)} \;\mathrm{d}(\alpha_1 + \alpha_2 + \alpha_3)(u)
    \\=& \int_{[0,\tau]} S_- \;\mathrm{d}\left(\alpha_2 + \frac{1}{2}\alpha_3\right) - \int_{[0,\tau]} \frac{\int_{(u,\tau]}S_-\;\mathrm{d}\left(A_2 + \frac{1}{2}A_3\right)}{1-\Delta A(u)} \;\mathrm{d}(\alpha_1 + \alpha_2 + \alpha_3)(u)
\end{align*}
 for all $\alpha_1,\alpha_2,\alpha_3\in D[0,\tau]$ by the chain rule, where we set $0/0 := 0$.

Furthermore,
Theorem~4.1 in \cite{dobler2017} provides that 
\begin{align}\label{eq:merken}
    \sqrt{n} \left( \widehat{A}_{1} - {A}_{1}, \widehat{A}_{2} - {A}_{2}, \widehat{A}_{3} - {A}_{3} \right) \xrightarrow{d} \left(U_{1},U_2,U_3  \right)
\end{align} holds as $n\to\infty$ on $D^3[0,\tau]$, 
where $U_{1}, U_2, U_3$ are zero-mean Gaussian-martingales with
\begin{align*}
    &Cov (U_{j}(t), U_{j}(s)) = \int_{[0,{t \wedge s}]} \frac{1 - \Delta {A}_{j}}{y} \;\mathrm{d}{A}_{j} =: {\sigma}_{j}^2(t \wedge s)
   ,\\& Cov (U_{j}(t), U_{\ell}(s)) = - \int_{[0,{t \wedge s}]} \frac{\Delta {A}_{\ell}}{y} \;\mathrm{d}{A}_{j} =: {\sigma}_{j\ell}(t\wedge s) 
\end{align*}
with $y(t) := S_-(t)G_-(t)$ for all $t,s\in [0,\tau], j,\ell\in\{1,2,3\},j \neq \ell$. 
 By \cite{Erratum}, the limit variable is separable.

Thus, the delta-method (Theorem~3.10.4 in \cite{vaartWellner2023}) implies $\sqrt{n}(\widehat {\theta} - {\theta})\xrightarrow{d} \Psi'_{(A_1,A_2,A_3)}(U_1,U_2,U_3)$ as $n\to\infty$, where $\Psi'_{(A_1,A_2,A_3)}(U_1,U_2,U_3)$ follows a centered normal distribution.
The variance of $\Psi'_{(A_1,A_2,A_3)}(U_1,U_2,U_3)$ can be calculated as
\begin{align*}
    \sigma_\theta^2 &:= %\Var\left( \Psi'_{(A_1,A_2,A_3)}(U_1,U_2,U_3) \right)\\&=
    Var\left( \int_{[0,\tau]} S_- \;\mathrm{d}\left(U_2 + \frac{1}{2}U_3\right) - \int_{[0,\tau]} \frac{\int_{(u,\tau]}S_-\;\mathrm{d}\left(A_2 + \frac{1}{2}A_3\right)}{1-\Delta A(u)} \;\mathrm{d}(U_1 + U_2 + U_3)(u) \right)
    \\&=\int_{[0,\tau]} S_-^2 \;\mathrm{d}\left(\sigma^2_2 + \sigma_{23} + \frac{1}{4}\sigma^2_3\right) \\ &\quad -2 \int_{[0,\tau]} \frac{S_-(u)}{1-\Delta A(u)}\int_{(u,\tau]}S_-\;\mathrm{d}\left(A_2 + \frac{1}{2}A_3\right) \;\mathrm{d}\left(\sigma_{12} + \sigma_2^2 + \frac{3}{2}\sigma_{23} + \frac{1}{2}\sigma_{13} + \frac{1}{2}\sigma_3^2\right)(u)
    \\ &\quad + \int_{[0,\tau]} \frac{\left(\int_{(u,\tau]}S_-\;\mathrm{d}\left(A_2 + \frac{1}{2}A_3\right)\right)^2}{(1-\Delta A(u))^2} \;\mathrm{d}\sigma_{\bullet}^2(u)
    \\&= \int_{[0,\tau]} S_-^2 \;\mathrm{d}\left(\sigma^2_2 + \sigma_{23} + \frac{1}{4}\sigma^2_3\right) \\ &\quad -2 \int_{[0,\tau]}\int_{[0,v)}\frac{ S_-(u)S_-(v)}{1-\Delta A(u)}\;\mathrm{d}\left(\sigma_{12} + \sigma_2^2 + \frac{3}{2}\sigma_{23} + \frac{1}{2}\sigma_{13} + \frac{1}{2}\sigma_3^2\right)(u)\mathrm{d}\left(A_2 + \frac{1}{2}A_3\right)(v)
    \\ &\quad + \int_{[0,\tau]}\int_{[0,\tau]} S_-(u)S_-(v) \int_{[0,\min\{u,v\})} \frac{1}{(1-\Delta A(w))^2}\;\mathrm{d}\sigma_{\bullet}^2(w)\mathrm{d}\left(A_2 + \frac{1}{2}A_3\right)(u)\mathrm{d}\left(A_2 + \frac{1}{2}A_3\right)(v)
\end{align*}
 with $\sigma^2_\bullet := \sigma_1^2 + \sigma_2^2 + \sigma_3^2 + 2\sigma_{12} + 2\sigma_{13} + 2\sigma_{23}$.  
\hfill\qed

\subsection{Proof of Theorem~\ref{p_asymptotic}}

\begin{lem}\label{pdPaired}
    Under at least one of (1)--(3) in Assumption~\ref{assPaired}, we have $\sigma^2_\theta > 0$.
\end{lem}
\begin{proof}[Proof of Lemma~\ref{pdPaired}]
By the proof of Theorem~\ref{Normality}, it holds that
\begin{align*}
    \sigma^2_\theta &=  Var \left( \sum_{m=1}^3 \int\limits_{[0,\tau]} 
    h_{m} \;\mathrm{d}U_{m} \right)
\end{align*}
with \begin{align*}
    h_{1}(u) &:= \frac{\int_{(u,\tau]} S_-\;\mathrm{d}\left(A_2 + \frac{1}{2} A_3\right)}{1-\Delta A(u)},
    \\ h_{2}(u) &:= S_-(u) - \frac{\int_{(u,\tau]} S_-\;\mathrm{d}\left(A_2 + \frac{1}{2} A_3\right)}{1-\Delta A(u)}\\
    \text{and }\quad h_{3}(u) &:= \frac{S_-(u)}{2}- \frac{\int_{(u,\tau]} S_-\;\mathrm{d}\left(A_2 + \frac{1}{2} A_3\right)}{1-\Delta A(u)}
\end{align*} for all $u\in [0,\tau]$, where $0/0 := 0$. 
We can calculate this variance further as
\begin{align}
    \sigma^2_\theta &= \sum_{m=1}^3 \E\left(\left(\,\int\limits_{[0,\tau]} 
    h_{m} \;\mathrm{d}U_{m} \right)^2\right) + \sum_{m=1}^3\sum_{ \Tilde{m} \neq m} \E\left(\,\int\limits_{[0,\tau]} 
    h_{m} \;\mathrm{d}U_{m} \int\limits_{[0,\tau]} 
    h_{\Tilde{m}} \;\mathrm{d}U_{\Tilde{m}} \right)
    \nonumber\\&= \sum_{m=1}^3 \int\limits_{[0,\tau]} 
    h_{m}^2 \frac{1 - \Delta A_{m}}{y} \;\mathrm{d}A_{m}  - \sum_{m=1}^3\sum_{ \Tilde{m} \neq m} \int\limits_{[0,\tau]} 
    h_{m}h_{\Tilde{m}} \frac{\Delta A_{m}}{y}\;\mathrm{d}A_{\Tilde{m}}
    \nonumber\\&= \sum_{m=1}^3 \int\limits_{[0,\tau]} 
    \frac{h_{m}^2}{y} \;\mathrm{d}A_{m}  - \sum_{m=1}^3\sum_{ \Tilde{m} =1}^3 \int\limits_{[0,\tau]} 
    h_{m}h_{\Tilde{m}} \frac{\Delta A_{m}}{y}\;\mathrm{d}A_{\Tilde{m}}
    \nonumber\\&= \sum_{m=1}^3 \int\limits_{[0,\tau]} 
    \frac{h_{m}^2}{y} \;\mathrm{d}A^c_{m} + \sum_{x\in\mathcal{D}} \frac{\sum_{m=1}^3 h_{m}^2(x) \Delta A_{m}(x)
    - 
   \left(\sum_{m=1}^3 h_{m}(x) {\Delta A_{m}}(x) \right)^2}{y(x)} \label{eq:here2}
\end{align}
where $\mathcal{D} = \{x\in [0,\tau] : \Delta A(x)  > 0 \}$ is the set of discontinuity time points and $$A_{m}^c(x) := A_{m}(x) - \sum\limits_{y\leq x, y\in\mathcal{D}}\Delta A_{m}(y), m\in\{1,2,3\},$$ denotes the continuous part of $A_{m}$ at $x\in[0,\tau]$.
%\Dennisinline{Um hier den continuous part zu definieren, muesstest du die Zuwaechse der diskreten Komponente kumuliert abziehen, gell?}\Merleinline{Ja, habs korrigiert}
 The Cauchy-Schwarz inequality yields
 \begin{align*}
     \left(\sum_{m=1}^3 h_{m}(x) {\Delta A_{m}}(x) \right)^2 \leq \left(\sum_{m=1}^3 h_{m}^2(x) {\Delta A_{m}}(x)\right)\left( \sum_{m=1}^3 {\Delta A_{m}}(x) \right)% = \sum_{m=1}^3 h_{m}^2(x) {\Delta A_{m}}(x)  {\Delta A_{i}}(x) 
 \end{align*} and, thus, 
 \begin{align*}
     \sum_{m=1}^3 h_{m}^2(x) \Delta A_{m}(x)    -    \left(\sum_{m=1}^3 h_{m}(x) {\Delta A_{m}}(x) \right)^2
     & \geq %\sum_{m=1}^3 h_{m}^2(x) \Delta A_{m}(x)    -    \sum_{m=1}^3 h_{m}^2(x) {\Delta A_{m}}(x) \sum_{m=1}^3 {\Delta A_{m}}(x) \\&= 
     \sum_{m=1}^3 h_{m}^2(x) \Delta A_{m}(x) \left(1   -     {\Delta A}(x) \right) \\&\geq 0
 \end{align*} for all $x \in\mathcal{D}$, where $1-\Delta A(u) \geq S_{-}(\tau) = \P(\delta T_1 \geq\tau, T_2 \geq \tau) > 0$ due to Assumption~\ref{assPaired}.

 Under (1), we have $F_{1,-}(\tau) = \P(T_2 < \min\{\delta T_1, \tau\}) > 0$ and
 \begin{align*}
     h_1(u) &\geq \frac{S_-(\tau) \Delta (A_2 + \frac{1}{2}A_3) (\tau)}{1-\Delta A(u)} \geq \frac{\frac{1}{2}\Delta (F_2 + F_3) (\tau)}{1-\Delta A(u)} \\&= \frac{\frac{1}{2} \P(\min\{\delta T_1,\delta \tau_1\} \leq \min\{T_2,\tau_2\}, \delta T_1 \geq \tau, T_2 \geq \tau)}{1-\Delta A(u)} > 0
 \end{align*} for all $u\in[0,\tau)$.
 Hence, at least one of the summands in (\ref{eq:here2}) with $m=1$ is strictly positive.
 
Similarly, under (2), we have $F_{2,-}(\tau) = \P(\delta T_1 < \min\{T_2, \tau\}) > 0$ and
 \begin{align*}
     h_2(u) &= \frac{S(u) - F_2(\tau) + F_2(u) - \frac{1}{2}(F_3(\tau) - F_3(u))}{1-\Delta A(u)}
     \\&= \frac{F_1(\tau) - F_1(u) + \frac{1}{2}(F_3(\tau) - F_3(u))}{1-\Delta A(u)}
     \\&\geq \frac{ \frac{1}{2}\Delta (F_1 + F_3)(\tau) }{1-\Delta A(u)}
     \\&= \frac{\frac{1}{2} \P(\min\{\delta T_1,\delta \tau_1\} \geq \min\{T_2,\tau_2\}, \delta T_1 \geq \tau, T_2 \geq \tau)}{1-\Delta A(u)} > 0
 \end{align*} for all $u\in[0,\tau)$. Then, at least one of the summands in (\ref{eq:here2}) with $m=2$ is strictly positive.
 
For (3), we have $F_{3,-}(\tau) = \P(\delta T_1 = T_2 < \tau) > 0$ and
\begin{align*}
     |h_3(u)| &= \frac{\left|\frac{1}{2}S(u) - F_2(\tau) + F_2(u) - \frac{1}{2}(F_3(\tau) - F_3(u))\right|}{1-\Delta A(u)}
     \\&= \frac{\frac{1}{2}\left|F_1(\tau) - F_1(u) - F_2(\tau) + F_2(u)\right|}{1-\Delta A(u)}
     %\\&\geq \frac{ \Delta (F_1 + F_3)(\tau) }{2-2\Delta A(u)}
    % \\&= \frac{ \left|\P(\min\{\delta T_1,\delta \tau_1\} > \min\{T_2,\tau_2\}, \min\{\delta T_1, T_2\} > u) - \P(\min\{\delta T_1,\delta \tau_1\} < \min\{T_2,\tau_2\}, \min\{\delta T_1, T_2\} > u)\right|}{2-2\Delta A(u)}\\& 
    > 0
 \end{align*} for all $u\in[0,\tau)$. Thus, at least one of the summands in (\ref{eq:here2}) with $m=3$ is strictly positive.
 %Therefore, consider different cases. 
 %Case 1: $F_m(\tau), F_{\tilde{m}}(\tau) > 0$ for two distinct indices $m,\tilde{m}\in\{1,2,3\}, m\neq \tilde{m}$. Assume that $h_m(u) = h_{\tilde{m}}(u) = 0$ for some $u\in[0,\tau)$. Then, the definitions of $h_m, h_{\tilde{m}}$ imply $0 = S_-(u) \geq S_-(\tau) > 0$, which yields a contradiction. Hence, for all $u\in[0,\tau)$, we have $h_m(u) \neq 0$ or $h_{\tilde{m}}(u) \neq 0$. Since $F_m(\tau), F_{\tilde{m}}(\tau) > 0$
\end{proof}
%\begin{proof}[Proof of Theorem~\ref{p_asymptotic}]

 Analogously to \cite{dobler24}, we can use the Greenwood-type variance estimators $\widehat\sigma_m^2, \widehat\sigma_{m\ell}, \widehat\sigma^2_\bullet$ for $\sigma_m^2, \sigma_{m\ell}, \sigma^2_\bullet$, cf. \cite{andersen93}, (4.4.17) and (4.4.18), and $\widehat A = \widehat A_1 + \widehat A_2 + \widehat A_3$ for $A$ to obtain the variance estimator
 \begin{align}\begin{split}\label{eq:varestpaired}
    \widehat\sigma_\theta^2 &= \int_{[0,\tau]} \widehat S_-^2 \;\mathrm{d}\left(\widehat \sigma^2_2 + \widehat \sigma_{23} + \frac{1}{4}\widehat \sigma^2_3\right) \\ &\quad -2 \int_{[0,\tau]}\int_{[0,v)}\frac{ \widehat S_-(u)\widehat S_-(v)}{1-\Delta \widehat A(u)}\;\mathrm{d}\left(\widehat \sigma_{12} + \widehat \sigma_2^2 + \frac{3}{2}\widehat \sigma_{23} + \frac{1}{2}\widehat \sigma_{13} + \frac{1}{2}\widehat \sigma_3^2\right)(u)\mathrm{d}\left(\widehat A_2 + \frac{1}{2}\widehat A_3\right)(v)
    \\ &\quad + \int_{[0,\tau]}\int_{[0,\tau]} \widehat S_-(u)\widehat S_-(v) \int_{[0,\min\{u,v\})} \frac{1}{(1-\Delta \widehat A)^2}\;\mathrm{d}\widehat \sigma_{\bullet}^2\mathrm{d}\left(\widehat A_2 + \frac{1}{2}\widehat A_3\right)(u)\mathrm{d}\left(\widehat A_2 + \frac{1}{2}\widehat A_3\right)(v).
\end{split}\end{align}
 The variance estimator $\widehat \sigma^2_{\theta}$ is a continuous functional of $(\widehat A_1,\widehat A_2,\widehat A_3)$ and $\widehat\sigma_j^2, \widehat\sigma_{j\ell}, \widehat\sigma^2_\bullet, j,\ell\in\{1,2,3\}, j\neq \ell$. By the uniform consistency of $(\widehat A_1,\widehat A_2,\widehat A_3)$, cf. \eqref{eq:merken}, and $\widehat\sigma_j^2$, $\widehat\sigma_{j\ell}$, $\widehat\sigma^2_\bullet$, $ j,\ell\in\{1,2,3\}, j\neq \ell$, the continuous mapping theorem yields the consistency of the variance estimator. The theorem follows then by applying Slutzky's lemma.
%\end{proof}
\hfill\qed

\subsection{Proof of Theorem~\ref{p_resampling}}\label{ssec:proofp_resampling}
Analogously to the proof of Theorem~2 in the supplement of \cite{dobler24}, we obtain by Theorem~2 in \cite{dobler_bernoulli} that $\sqrt{n}(\widetilde{\widehat\theta} - 1/2)$ converges weakly conditionally on the data $(Z_j,\varepsilon_j), j\in\{1,\dots,n\},$ in outer probability as $n\to\infty$ to a centered normal variable with variance
\begin{align*}
   \widetilde\sigma_\theta^2 &:=   \int_{[0,\tau]} S_-^2 \;\mathrm{d}\left(\widetilde\sigma^2_2 + \widetilde\sigma_{23} + \frac{1}{4}\widetilde\sigma^2_3\right) \\ &\quad -\int_{[0,\tau]}\int_{[0,v)}\frac{ S_-(u)S_-(v)}{1-\Delta A(u)}\;\mathrm{d}\left(\widetilde\sigma_{12} + \widetilde\sigma_2^2 + \frac{3}{2}\widetilde\sigma_{23} + \frac{1}{2}\widetilde\sigma_{13} + \frac{1}{2}\widetilde\sigma_3^2\right)(u)\mathrm{d}A(v)
    \\ &\quad + \frac{1}{4}\int_{[0,\tau]}\int_{[0,\tau]} S_-(u)S_-(v) \int_{[0,\min\{u,v\})} \frac{1}{(1-\Delta A(w))^2}\;\mathrm{d}\widetilde\sigma_{\bullet}^2(w)\mathrm{d}A(u)\mathrm{d}A(v),
\end{align*} where \begin{align*}
    &\widetilde{\sigma}_{1}^2(t) := \widetilde{\sigma}_{2}^2(t) := \frac{1}{2}\int_{[0,{t}]} \frac{1 - \Delta \frac{A_1+A_2}{2}}{y} \;\mathrm{d}(A_1+A_2)
    ,\quad \widetilde{\sigma}_{3}^2(t) := \sigma_3^2(t)
   ,\\& \widetilde{\sigma}_{12}(t) := - \frac{1}{4}\int_{[0,{t}]} \frac{\Delta (A_1+A_2)}{y} \;\mathrm{d}(A_1+A_2)
   ,\quad \widetilde{\sigma}_{13}(t) := \widetilde{\sigma}_{23}(t) := - \frac{1}{2}\int_{[0,{t}]} \frac{\Delta A_3}{y} \;\mathrm{d}({A}_{1}+A_2),\\
   &\tilde{\sigma}^2_{\bullet} := \tilde{\sigma}_1^2 + \tilde{\sigma}_2^2 + \tilde{\sigma}_3^2 + 2\tilde{\sigma}_{12}+ 2\tilde{\sigma}_{13}+ 2\tilde{\sigma}_{23}
\end{align*}
for all $t\in [0,\tau]$.

\begin{lem}\label{pdPairedresampling}
    If Assumption~\ref{assPaired1} and $\max\{\P(T_2 < \min\{\delta T_1, \tau\}), \P(\delta T_1 < \min\{T_2, \tau\})\} >0$ hold, we have $\widetilde\sigma_\theta^2 > 0$.
\end{lem}
\begin{proof}[Proof of Lemma~\ref{pdPairedresampling}]
    By proceeding similarly as in the proof of Lemma~\ref{pdPaired}, we obtain
    \begin{align*}
   \widetilde\sigma_\theta^2 &=   \frac{1}{2} \left(\sum_{m=1}^2 \int\limits_{[0,\tau]} 
    \frac{\widetilde h_{m}^2}{y} \;\mathrm{d}(A_1^c+A_2^c)\sum_{x\in\mathcal{D}} \frac{\sum_{m=1}^2 \widetilde h_{m}^2(x) \Delta (A_1+A_2)(x)
    - 
   \left(\sum_{m=1}^2 \widetilde h_{m}(x) {\Delta (A_1+A_2)}(x) \right)^2}{y(x)}\right),
\end{align*}
where \begin{align*}
    \widetilde h_{1}(u) &:= \frac{\int_{(u,\tau]} S_-\;\mathrm{d}A}{2-2\Delta A(u)}
    \quad\text{ and }\quad \widetilde h_{2}(u) := S_-(u) - \widetilde h_{1}(u)
\end{align*} for all $u\in [0,\tau]$ with $0/0 := 0$.
As in the proof of Lemma~\ref{pdPaired}, one can show
\begin{align*}
    \widetilde h_{1}(u) \geq \frac{S_-(\tau)\Delta A(\tau)}{2-2\Delta A(u)} = \frac{\Delta (F_1+F_2+F_3)(\tau)}{2-2\Delta A(u)}= \frac{\P(\delta T_1 \geq \tau, T_2 \geq \tau)}{2-2\Delta A(u)} > 0
\end{align*} for all $u\in[0,\tau)$ by Assumption~\ref{assPaired1}. Furthermore, it holds 
\begin{align*}
    F_{1,-}(\tau) + F_{2,-}(\tau) &= \P(T_2 < \min\{\delta T_1,\tau\} \lor \delta T_1 < \min\{T_2,\tau\})\\&\geq \max\{\P(T_2 < \min\{\delta T_1, \tau\}), \P(\delta T_1 < \min\{T_2, \tau\})\} > 0.
\end{align*}
Hence, at least one of the summands with $m=1$ is strictly positive. 
\end{proof}

The consistency of the variance estimator $\widetilde{\widehat\sigma}_{\theta}^2$ for $\widetilde\sigma_\theta^2$ follows analogously as in the proof of Theorem~\ref{p_asymptotic}.
Therefore note that $S, G$ and $A_3$ as well as their estimators remain the same for the randomized data $(Z_i,\tilde\varepsilon_i),i\in\{1,\dots,n\}.$ 
Moreover, $\frac{A_1+A_2}{2}, \frac{A_1+A_2}{2}, A_3$ can be calculated as the cause-specific cumulative hazard functions of the randomized data $(Z_j,\tilde\varepsilon_j),j\in\{1,\dots,n\}.$
Hence, the cause-specific Nelson-Aalen estimators based on the randomized data converge uniformly in outer probability to $\frac{A_1+A_2}{2}, \frac{A_1+A_2}{2}, A_3$, respectively, on $[0,\tau]$ by Theorem~4.1 in \cite{dobler2017} and the separability of the limit by \cite{Erratum}.
This implies the consistency of the variance estimator.

Applying Slutzky's lemma completes the proof.
\hfill\qed

\subsection{Proof of Theorem~\ref{rmst_asymptotic}}\label{ssec:proof3}
Let $G_j: [0,\infty)\ni t \mapsto \P ( C_j > t) \in [0,1]$ denote the survival function of the censoring times $C_j$ at time $t\geq 0$ in the following for $j\in\{1,2\}$.
By Appendix~D.1 of \cite{dobler_bernoulli}, the influence function of the Kaplan-Meier estimator $\widehat S_i$ at the Dirac measure in ${(Y_{1i}, Y_{2i},\Delta_{1i},\Delta_{2i})}$ is given by
$$
t \mapsto S_j(t)\left[\frac{\Delta_{ji}{1}\{Y_{ji} \leq t\}}{G_{j-}(Y_{ji}) S_{j}(Y_{ji})} - \int_{[0,\min\{t,Y_{ji}\}]} \frac{1}{G_{j-}(u)S_{j}(u)}\;\mathrm{d}A_j(u) \right]
$$ for $j\in\{1,2\}, i\in\{1,...,n\}$.
Hence, one can calculate the influence function of the estimator $\widehat\mu_j$ for the RMST $\mu_j$ at the Dirac measure in ${(Y_{1i}, Y_{2i},\Delta_{1i},\Delta_{2i})}$ as
\begin{align*}
   \mathrm{IF}_{j}(Y_{ji},\Delta_{ji}) &:= \int_0^{\tau} S_j(t)\left[\frac{\Delta_{ji}{1}\{Y_{ji} \leq t\}}{G_{j-}(Y_{ji}) S_{j}(Y_{ji})} - \int_{[0,\min\{t,Y_{ji}\}]} \frac{1}{G_{j-}(u)S_{j}(u)}\;\mathrm{d}A_j(u) \right] \;\mathrm{d}t
   % kann man noch weiter umformen wenn man möchte, siehe main.pdf von Marc
\end{align*}
for $j\in\{1,2\}, i\in\{1,...,n\}$. Furthermore, we set
\begin{align*}
   \mathrm{IF}_{j}(x,d) &:= \int_0^{\tau} S_j(t)\left[\frac{d {1}\{x \leq t\}}{G_{j-}(x) S_{j}(x)} - \int_{[0,\min\{t,x\}]} \frac{1}{G_{j-}(u)S_{j}(u)}\;\mathrm{d}A_j(u) \right] \;\mathrm{d}t
   % kann man noch weiter umformen wenn man möchte, siehe main.pdf von Marc
\end{align*}
for all $j\in\{1,2\}, x\in[0,\tau],d\in\{0,1\}$.
Note that $\sup_{x\in[0,\tau],d\in\{0,1\}} |\mathrm{IF}_{1}(x,d)|$ and $\sup_{x\in[0,\tau],d\in\{0,1\}} |\mathrm{IF}_{2}(x,d)|$ are bounded due to Assumption~\ref{ass_rmst}.
The influence function of $\widehat\mu_2 - \widehat\mu_1$ at the Dirac measure in $(Y_{1i}, Y_{2i},\Delta_{1i},\Delta_{2i})$ is then given by
\begin{align*}
    \mathrm{IF}^{\text{diff}}_i := \mathrm{IF}_{2}(Y_{2i},\Delta_{2i}) - \mathrm{IF}_{1}(Y_{1i},\Delta_{1i})
\end{align*}
and the influence function of $\log(\widehat\mu_2) - \log(\widehat\mu_1)$ at the Dirac measure in $(Y_{1i}, Y_{2i},\Delta_{1i},\Delta_{2i})$ is given by
\begin{align*}
    \mathrm{IF}^{\text{rat}}_i := \frac{\mathrm{IF}_{2}(Y_{2i},\Delta_{2i})}{\mu_2} - \frac{\mathrm{IF}_{1}(Y_{1i},\Delta_{1i})}{\mu_1}
\end{align*}
for $i\in\{1,...,n\}.$ 
By \cite{dobler_bernoulli}, it holds
\begin{align*}
    \sqrt{n}((\widehat\mu_2 - \widehat\mu_1) - (\mu_2 - \mu_1)) = \frac{1}{\sqrt{n}} \sum\limits_{i=1}^n \mathrm{IF}^{\text{diff}}_i + o_p(1)
\end{align*}
and
\begin{align*}
    \sqrt{n}((\log(\widehat\mu_2) - \log(\widehat\mu_1)) - (\log(\mu_2) - \log(\mu_1))) = \frac{1}{\sqrt{n}} \sum\limits_{i=1}^n \mathrm{IF}^{\text{rat}}_i + o_p(1)
\end{align*}
as $n\to\infty$. The advantage of this representation is that the summands are independent and identically distributed. Hence, the asymptotic normality follows by an application of the central limit theorem.

Now, it remains to show that we have consistent estimators $\widehat\sigma^2_{\text{diff}}, \widehat\sigma^2_{\text{rat}}$ for the limit variances $\sigma^2_{\text{diff}} := \mathbb Var(\mathrm{IF}^{\text{diff}}_1)$ and $\sigma^2_{\text{rat}}:= \mathbb Var(\mathrm{IF}^{\text{rat}}_1)$, respectively. Since  $\sup_{x\in[0,\tau],d\in\{0,1\}} |\mathrm{IF}_{1}(x,d)|$ and $\sup_{x\in[0,\tau],d\in\{0,1\}} |\mathrm{IF}_{2}(x,d)|$ are bounded, the variances exist. We define
$$ \widehat{\mathrm{IF}}_{j}(x,d) := \int_0^{\tau} \widehat S_j(t)\left[\frac{d\cdot {1}\{x \leq t\}}{\widehat G_{j-}(x)\widehat S_{j}(x)} - \int_{[0,\min\{t,x\}]} \frac{1}{\widehat G_{j-}(u)\widehat S_{j}(u)}\;\mathrm{d}\widehat A_j(u) \right] \;\mathrm{d}t$$
for $j\in\{1,2\}, x\in [0,\tau], d\in\{0,1\}$, where $\widehat S_j, \widehat G_j$ denote the Kaplan-Meier estimators of $S_i,G_i$, respectively, %$\widehat y_j(t) := n^{-1}\sum_{i=1}^n {1}\{Y_{ji}\geq t\}$ for all $t\geq 0$ 
and $\widehat A_j$ denotes the Nelson-Aalen estimator of $ A_j$.
Then, we set $\widehat{\mathrm{IF}}_i^{\text{diff}} := \widehat{\mathrm{IF}}_{2}(Y_{2i},\Delta_{2i}) - \widehat{\mathrm{IF}}_{1}(Y_{1i},\Delta_{1i}) $ and $\widehat{\mathrm{IF}}_i^{\text{rat}} := \widehat{\mathrm{IF}}_{2}(Y_{2i},\Delta_{2i})/\widehat\mu_2 - \widehat{\mathrm{IF}}_{1}(Y_{1i},\Delta_{1i})/\widehat\mu_1 $ for all $i\in\{1,...,n\}$. Consequently, $\widehat\sigma^2_{\text{diff}}, \widehat\sigma^2_{\text{rat}}$ can be estimated as the empirical variances
\begin{align*}
    \widehat\sigma^2_{\text{diff}} := \frac{1}{n} \sum_{i=1}^n \left(\widehat{\mathrm{IF}}_i^{\text{diff}} - \frac{1}{n}\sum_{j=1}^n\widehat{\mathrm{IF}}_j^{\text{diff}} \right)^2 
    \text{ and } \widehat\sigma^2_{\text{rat}} := \frac{1}{n} \sum_{i=1}^n \left(\widehat{\mathrm{IF}}_i^{\text{rat}} - \frac{1}{n}\sum_{j=1}^n\widehat{\mathrm{IF}}_j^{\text{rat}} \right)^2.
\end{align*}
It is well known that $\widehat S_j, \widehat G_j, \widehat  A_j$ are uniformly consistent for $S_j, G_j,  A_j$ on $[0,\tau]$, respectively, for $j\in\{1,2\}$, see, e.g., the supplement of \cite{RMST} for details.
Due to the continuity of the functionals, the continuous mapping theorem implies
$$ \sup\limits_{x\in[0,\tau],d\in\{0,1\}} \left|\widehat{\mathrm{IF}}_j(x,d) - {\mathrm{IF}}_j(x,d)  \right| \xrightarrow{\P} 0$$
%and 
%$$ \sup\limits_{y\in[0,\tau],d\in\{0,1\}} \left|\widehat{\mathrm{IF}}_j^2(y,d) - {\mathrm{IF}}_j^2(y,d)  \right| \xrightarrow{P} 0$$ 
as $n\to\infty$ for $j\in\{1,2\}$. Thus, easy calculations and an application of Slutzky's lemma yield
%\begin{align*}    \left| \widehat{\mathrm{IF}}_i^{\text{diff}} - {\mathrm{IF}}_i^{\text{diff}} \right| \leq \sup\limits_{y\in[0,\tau],d\in\{0,1\}} \left|\widehat{\mathrm{IF}}_1(y,d) - {\mathrm{IF}}_1(y,d)  \right| + \sup\limits_{y\in[0,\tau],d\in\{0,1\}} \left|\widehat{\mathrm{IF}}_2(y,d) - {\mathrm{IF}}_2(y,d)  \right| \xrightarrow{P} 0,\\    \left| (\widehat{\mathrm{IF}}_i^{\text{diff}})^2 - ({\mathrm{IF}}_i^{\text{diff}})^2 \right| =    \left| \widehat{\mathrm{IF}}_i^{\text{diff}} - {\mathrm{IF}}_i^{\text{diff}} \right|\left| \widehat{\mathrm{IF}}_i^{\text{diff}} + {\mathrm{IF}}_i^{\text{diff}}\right|    \leq \left| \widehat{\mathrm{IF}}_i^{\text{diff}} - {\mathrm{IF}}_i^{\text{diff}} \right| \left(\sup\limits_{y\in[0,\tau],d\in\{0,1\}} \left|\widehat{\mathrm{IF}}_1(y,d) - \widehat{\mathrm{IF}}_2(y,d)  \right| + \sup\limits_{y\in[0,\tau],d\in\{0,1\}} \left|{\mathrm{IF}}_1(y,d) - {\mathrm{IF}}_2(y,d)\right|\right) \xrightarrow{P} 0\end{align*} as $n\to\infty$ and, analogously,
\begin{align*}
   & \left|  \widehat\sigma^2_{\text{diff}} - \frac{1}{n} \sum_{i=1}^n \left({\mathrm{IF}}_i^{\text{diff}} - \frac{1}{n}\sum_{j=1}^n{\mathrm{IF}}_j^{\text{diff}} \right)^2 \right| \xrightarrow{P} 0 
    \quad\text{ and }\quad \\&\left|  \widehat\sigma^2_{\text{rat}} - \frac{1}{n} \sum_{i=1}^n \left({\mathrm{IF}}_i^{\text{rat}} - \frac{1}{n}\sum_{j=1}^n{\mathrm{IF}}_j^{\text{rat}} \right)^2 \right| \xrightarrow{P} 0
\end{align*} as $n\to\infty$. Since ${\mathrm{IF}}_i^{\text{diff}}, i\in\{1,...,n\},$ are i.i.d., it follows
    $\widehat\sigma^2_{\text{diff}} \xrightarrow{P} \sigma^2_{\text{diff}} $ and, analogously, $\widehat\sigma^2_{\text{rat}} \xrightarrow{P} \sigma^2_{\text{rat}}$ as $n\to\infty$.
%\end{proof}
\hfill\qed

\subsection{Proof of Theorem~\ref{rmst_randomization}}\label{ssec:proof4}
We aim to apply Theorem~2 in \cite{dobler_bernoulli} and, hence, verifying the conditions in the following. Therefore, let
\begin{align*}
    \mathcal{F} &:= \left\{ (y_1,y_2,d_1,d_2) \mapsto d_j \cdot {1}\{y_j \leq t\}, (y_1,y_2,d_1,d_2) \mapsto  {1}\{y_j > t\}
    \mid t\in [0,\tau], j\in\{1,2\} \right\}
    \end{align*} and \begin{align*} \widetilde{\mathcal{F}} &:= \left\{ (y_1,y_2,d_1,d_2) \mapsto \frac{1}{2}\left( f(y_1,y_2,d_1,d_2) + f(y_2,y_1,d_2,d_1) \right)
    \mid f \in\mathcal{F} \right\}.
\end{align*}
Analogously as in the proof of Theorem~2 in \cite{dobler24}, $\mathcal{F}$ and $\widetilde{\mathcal{F}}$ are VC-classes. Consequently, the sets are $\mathbb P$- and $\widetilde{\mathbb P}$-Donsker and Glivenko-Cantelli classes with $\mathbb P := \P^{(Y_{11},Y_{21},\Delta_{11},\Delta_{21})}$ and $\widetilde{\mathbb P} := \P^{(Y_{11}^{\pi}, Y_{21}^{\pi}, \Delta_{11}^{\pi}, \Delta_{21}^{\pi})}$. Furthermore, $\mathbb P$ and $\widetilde{\mathbb P}$ have bounded supremum norms with respect to both sets $\mathcal{F}$ and $\widetilde{\mathcal{F}}$. Moreover, the Kaplan-Meier estimator is a Hadamard-differentiable functional as shown in Example~3.10.33 in \cite{vaartWellner2023}. Hence, the estimators for the RMST difference and ratio are also Hadamard-differentiable functionals, respectively, by the chain rule in Lemma~3.10.3 in \cite{vaartWellner2023}. Thus, Theorem~2 in \cite{dobler_bernoulli} provides that $ \sqrt{n}(\widehat \mu_2^{\pi} - \widehat \mu_1^{\pi})  $ and $\sqrt{n}\log(\widehat \mu_2^{\pi} / \widehat \mu_1^{\pi})$ are converging in distribution to centered normal distributions in outer probability conditionally on the data $(Y_{1i},Y_{2i},\Delta_{1i},\Delta_{2i})$, $i\in\{1,...,n\}$, as $n\to\infty$. For deriving the variances of the limit distributions, let $S^{\pi}:= \frac{1}{2}(S_1 + S_2), G^{\pi} := \frac{1}{2}(G_1 + G_2), F^{\pi}:= 1-S^\pi$ and $ A^{\pi}(.) := \int_{[0,.]} \frac{1}{S^{\pi}_-(t)} \:\mathrm{d} F^{\pi}(t) $ be the pooled survival, distribution and cumulative hazard functions. Note that permuting the data randomly leads to those survival, distribution and cumulative hazard functions for the permuted data.
Furthermore,
define $${\mathrm{Q.IF}}(x,d) := \int_0^{\tau} S^{\pi}(t)\left[\frac{d {1}\{x \leq t\}}{G^{\pi}_{-}(x) S^{\pi}(x)} - \int_{[0,\min\{t,x\}]} \frac{1}{G^{\pi}_{-}(u)S^{\pi}(u)}\;\mathrm{d} A^{\pi}(u) \right] \;\mathrm{d}t$$ 
for all $ x\in [0,\tau], d\in\{0,1\}.$
Note that $\sup_{x\in[0,\tau], d\in\{0,1\}} | {\mathrm{Q.IF}}(x,d) |$ is bounded under Assumption~\ref{ass_rmst}.
One can show that the variances of the normal distributions are ${\sigma}_{\text{diff}}^{\pi}:=\mathbb Var\left( {\mathrm{Q.IF}}^{\text{diff}}_1 \right)$ and ${\sigma}_{\text{rat}}^{\pi}:=\mathbb Var\left( {\mathrm{Q.IF}}^{\text{rat}}_1 \right)$, where
${\mathrm{Q.IF}}^{\text{diff}}_1 := {\mathrm{Q.IF}}(Y_{21},\Delta_{21}) - {\mathrm{Q.IF}}(Y_{11},\Delta_{11})$ and
${\mathrm{Q.IF}}^{\text{rat}}_1 := ({\mathrm{Q.IF}}(Y_{21},\Delta_{21}) - {\mathrm{Q.IF}}(Y_{11},\Delta_{11}))/\overline{\mu}$
with $\overline{\mu} := \frac{\mu_1+\mu_2}{2}$.
%\Todo{Nochmal nachrechnen}

Hence, it remains to show that the permutation counterparts of the variance estimators are consistent.
% \Todo{Noch sehr kurz...}
Since $\widehat{S}_j^{\pi}, \widehat{G}_j^{\pi}, \widehat{A}_j^{\pi}, j\in\{1,2\},$ are continuous functionals of the empirical process of $( Y_{1i}^{\pi},Y_{2i}^{\pi}, \Delta_{1i}^{\pi}, \Delta_{2i}^{\pi})$ and $\mathcal{F}$ is a Glivenko-Cantelli class, it follows that $\widehat{S}_j^{\pi}, \widehat{G}_j^{\pi}, \widehat{A}_j^{\pi}$ are uniformly consistent for $S^{\pi}, G^{\pi}$ and $A^{\pi}$ on $[0,\tau]$, respectively, for $j\in\{1,2\}$. Thus, we get
\begin{align*}
    \sup\limits_{x\in[0,\tau], d\in\{0,1\}} \left| {\mathrm{IF}}_j^{\pi}(x,d) - {\mathrm{Q.IF}}(x,d)\right| \xrightarrow{P} 0
\end{align*} as $n\to\infty$ for $j\in\{1,2\}$.
Then, the consistency of the variance estimators follow analogously as in the proof of Theorem~\ref{rmst_asymptotic}.
%\end{proof}
\hfill\qed

\section{Additional Simulation Results}\label{sec:appendix_simus}
\subsection{Results by sample size}
\begin{figure}[H]
    \centering
	\includegraphics[width=0.75\textwidth]{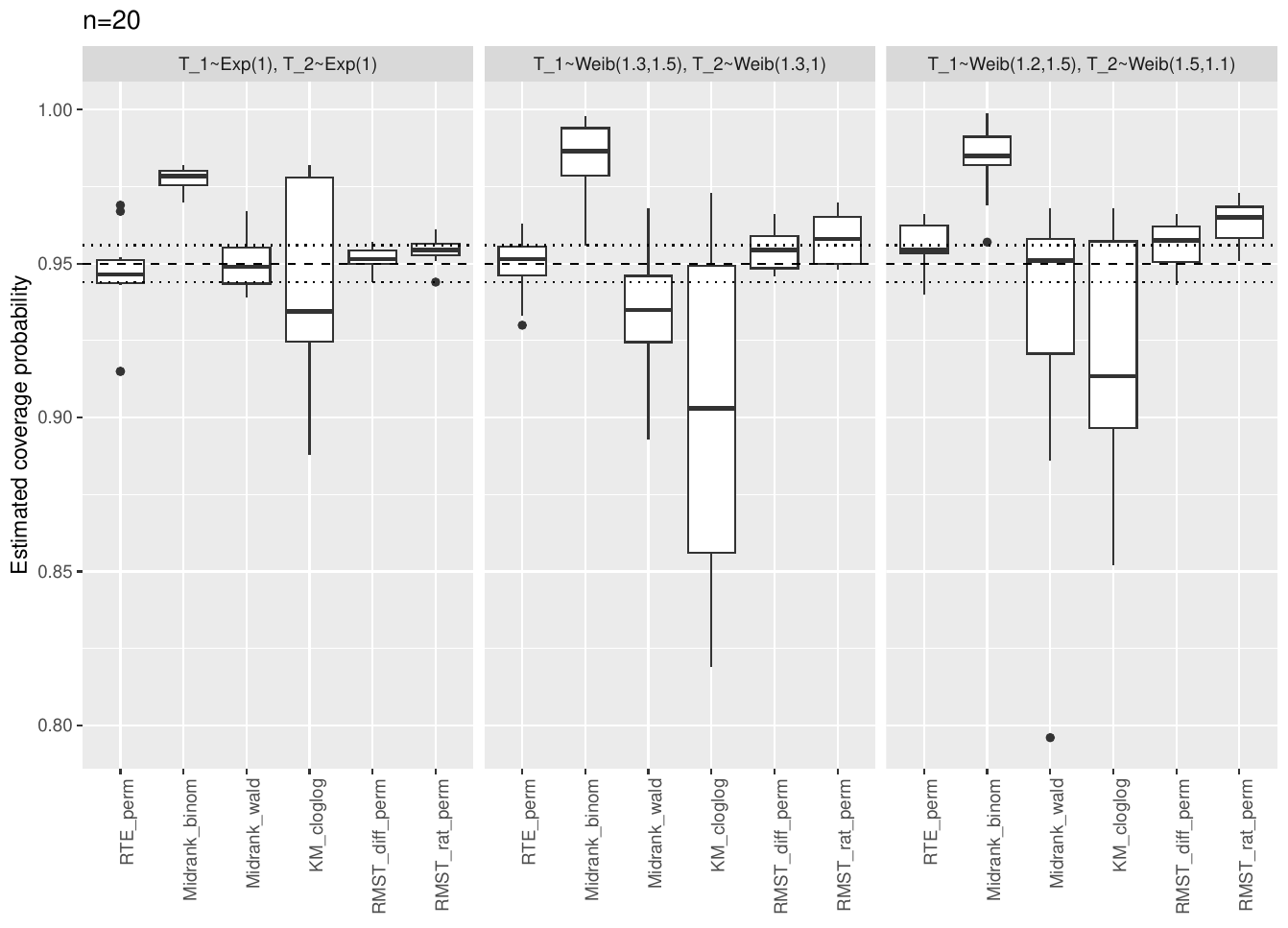}
	\caption{Estimated coverage probabilities for sample size $n=20$ of left-sided 95\% confidence intervals for $\delta=1$, stratified by marginal distribution. The dashed lines represent the borders of the binomial confidence interval $[94.4\%, 95.6\%]$.}
\label{fig:res_n20_left}
\end{figure}

\begin{figure}[H]
    \centering
	\includegraphics[width=0.75\textwidth]{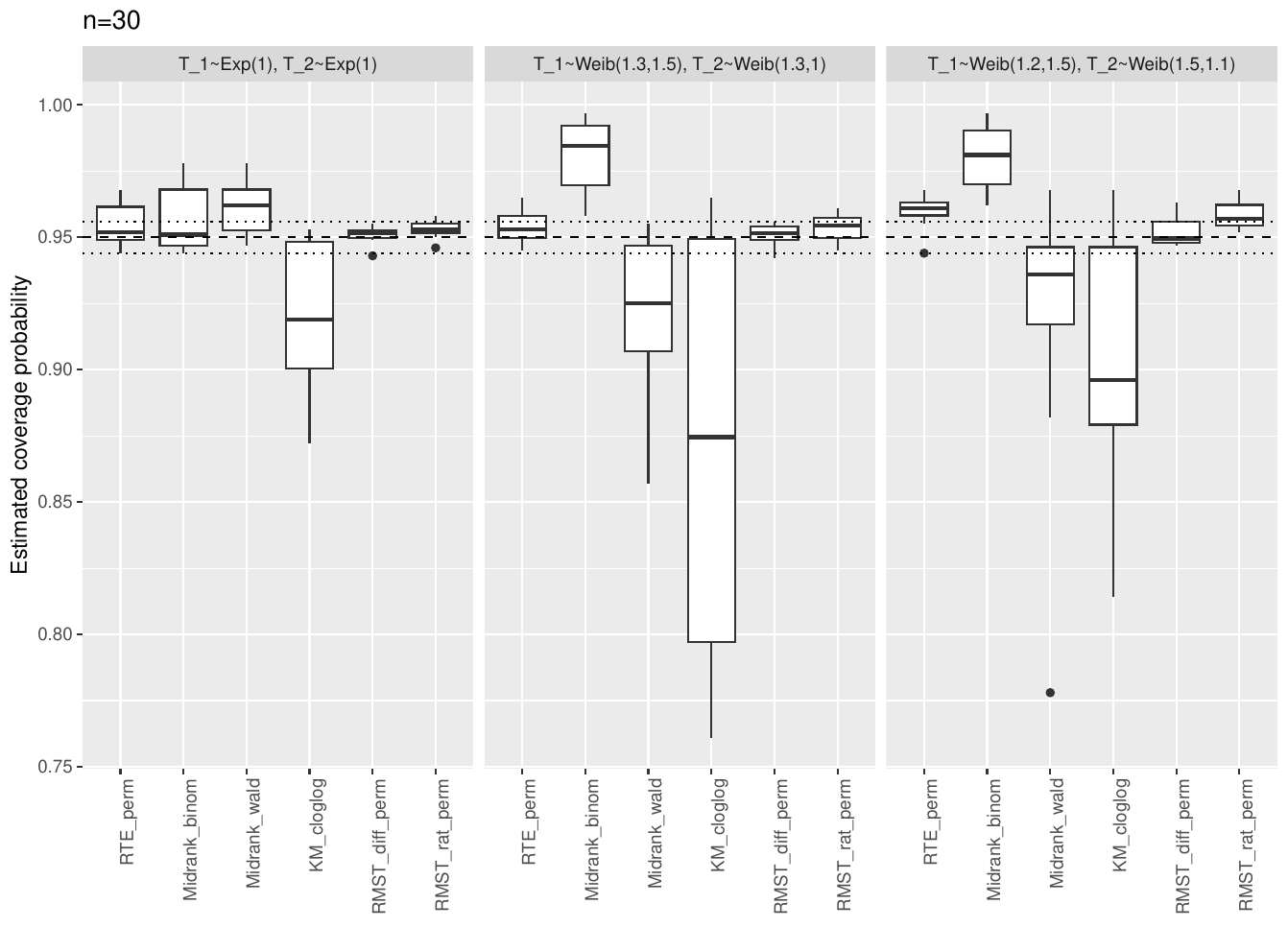}
	\caption{Estimated coverage probabilities for sample size $n=30$ of left-sided 95\% confidence intervals for $\delta=1$, stratified by marginal distribution. The dashed lines represent the borders of the binomial confidence interval $[94.4\%, 95.6\%]$.}
\label{fig:res_n30_left}
\end{figure}

\begin{figure}[H]
    \centering
	\includegraphics[width=0.75\textwidth]{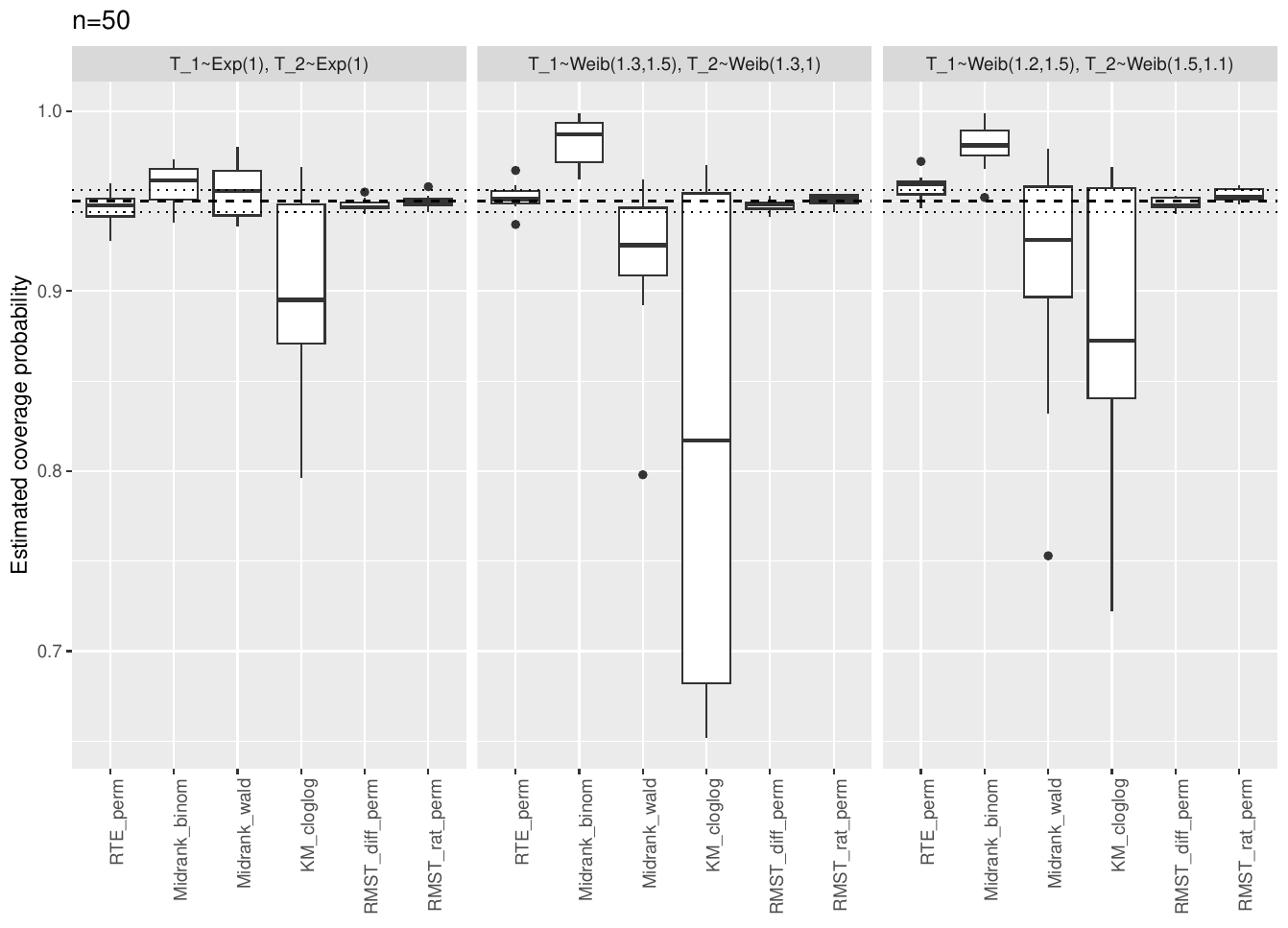}
	\caption{Estimated coverage probabilities for sample size $n=50$ of left-sided 95\% confidence intervals for $\delta=1$, stratified by marginal distribution. The dashed lines represent the borders of the binomial confidence interval $[94.4\%, 95.6\%]$.}
\label{fig:res_n50_left}
\end{figure}

\begin{figure}[H]
    \centering
	\includegraphics[width=0.75\textwidth]{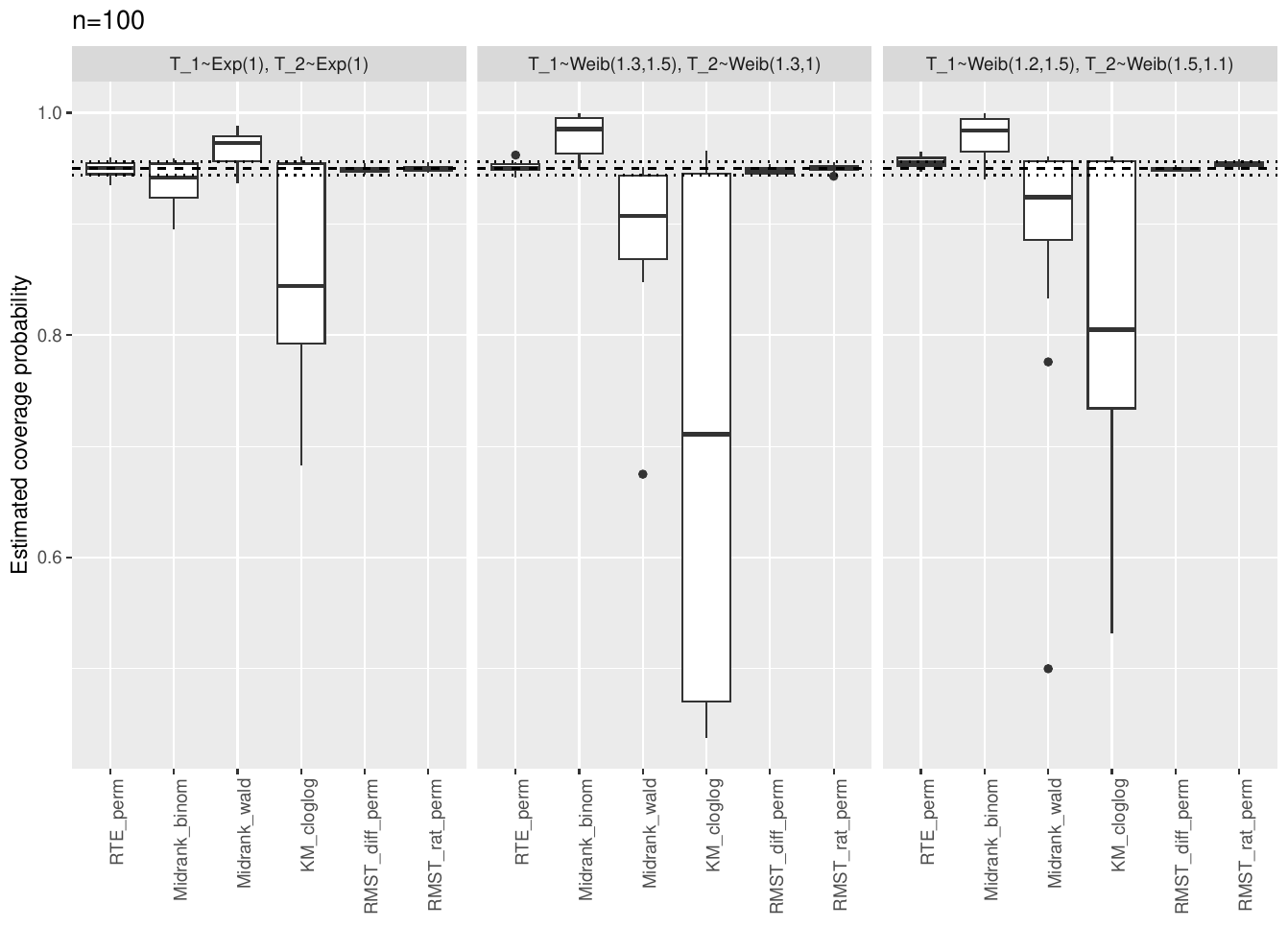}
	\caption{Estimated coverage probabilities for sample size $n=100$ of left-sided 95\% confidence intervals for $\delta=1$, stratified by marginal distribution. The dashed lines represent the borders of the binomial confidence interval $[94.4\%, 95.6\%]$.}
\label{fig:res_n100_left}
\end{figure}

\begin{figure}[H]
    \centering
	\includegraphics[width=0.75\textwidth]{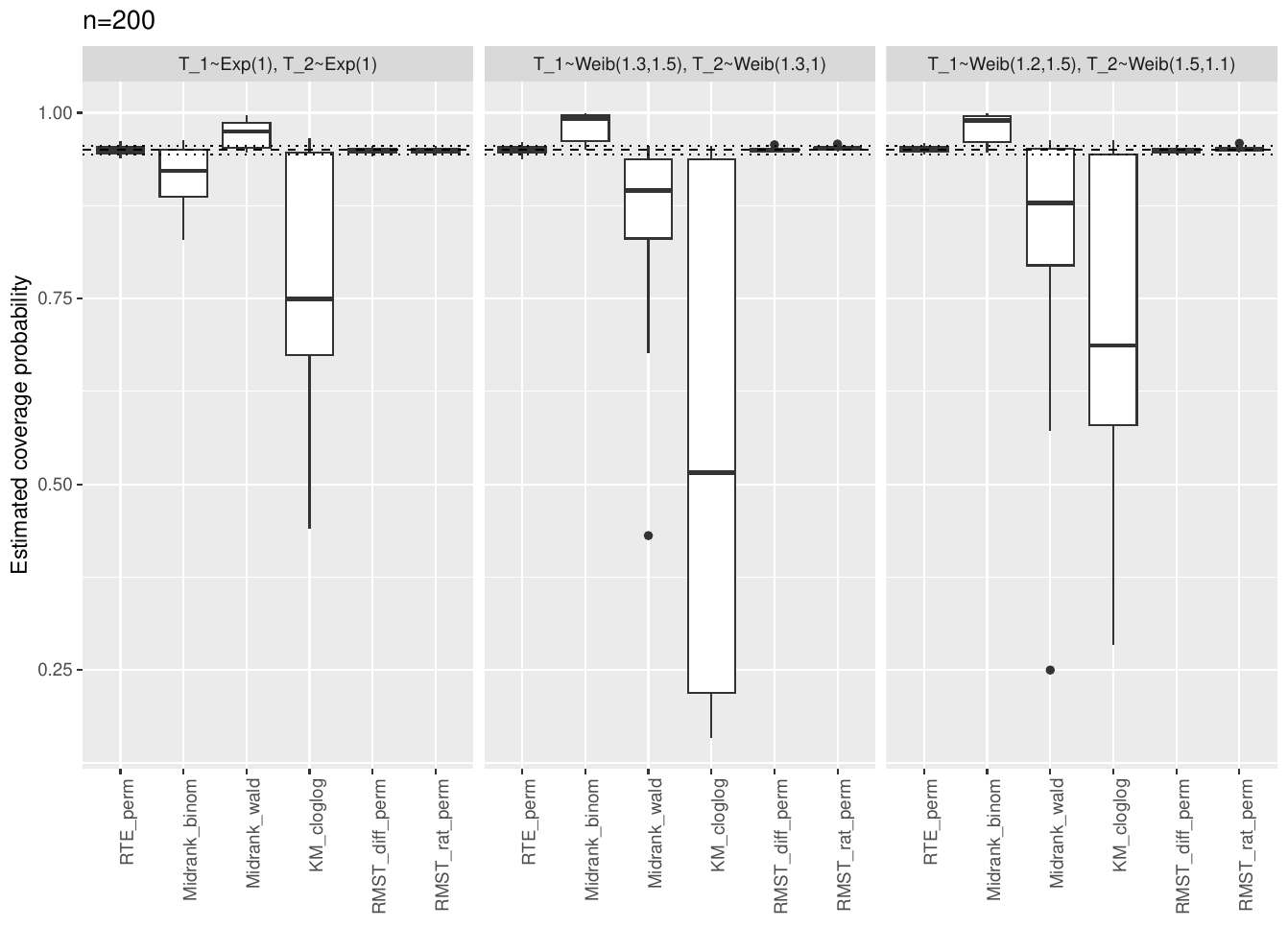}
	\caption{Estimated coverage probabilities for sample size $n=200$ of left-sided 95\% confidence intervals for $\delta=1$, stratified by marginal distribution. The dashed lines represent the borders of the binomial confidence interval $[94.4\%, 95.6\%]$.}
\label{fig:res_n200_left}
\end{figure}

\subsection{Results by dependence parameter}

\begin{figure}[H]
    \centering
	\includegraphics[width=0.75\textwidth]{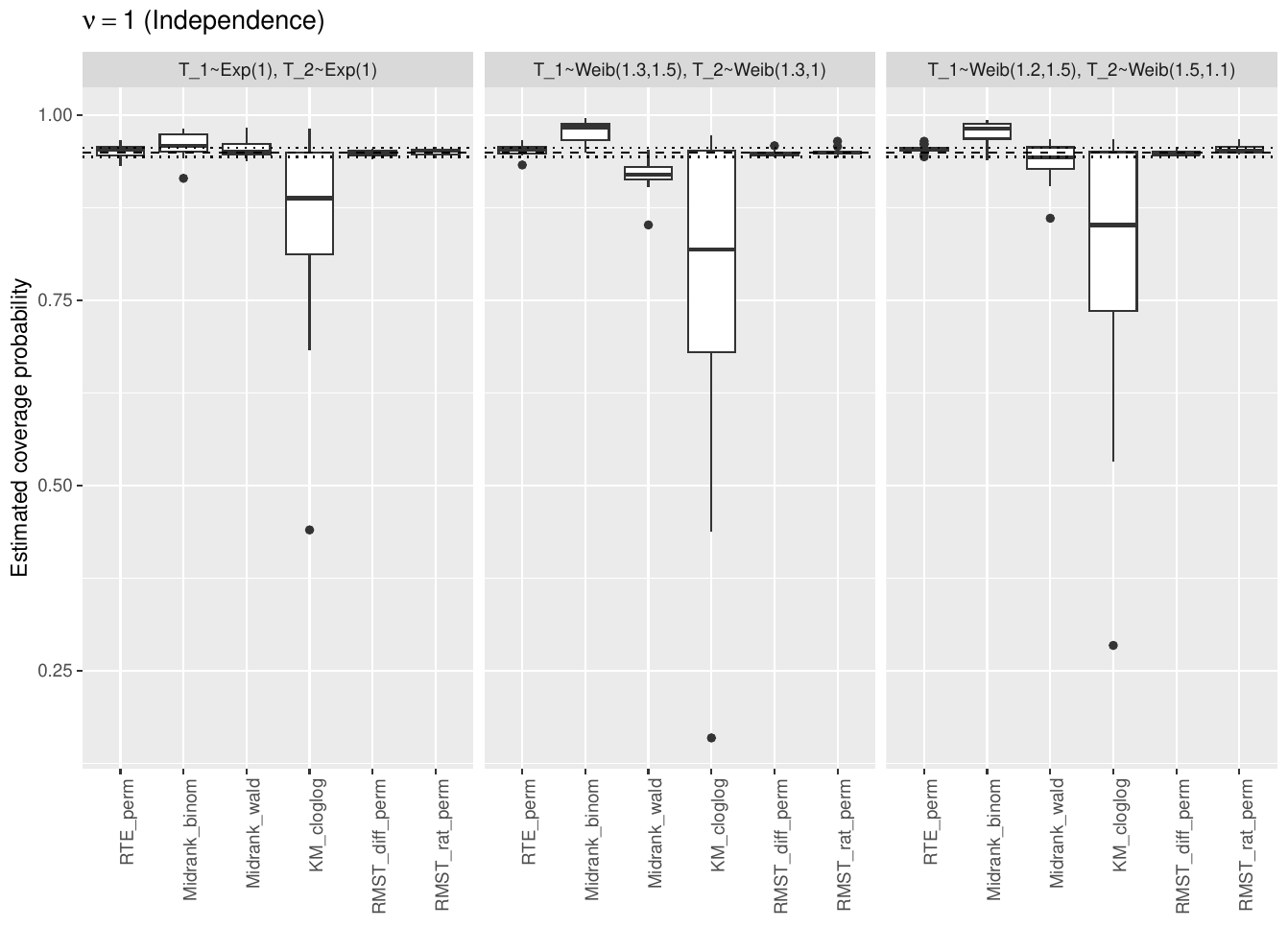}
	\caption{Estimated coverage probabilities for dependence parameter $\nu=1$ (independence) of left-sided 95\% confidence intervals for $\delta=1$, stratified by marginal distribution. The dashed lines represent the borders of the binomial confidence interval $[94.4\%, 95.6\%]$.}
\label{fig:res_nu1_left}
\end{figure}

\begin{figure}[H]
    \centering
	\includegraphics[width=0.75\textwidth]{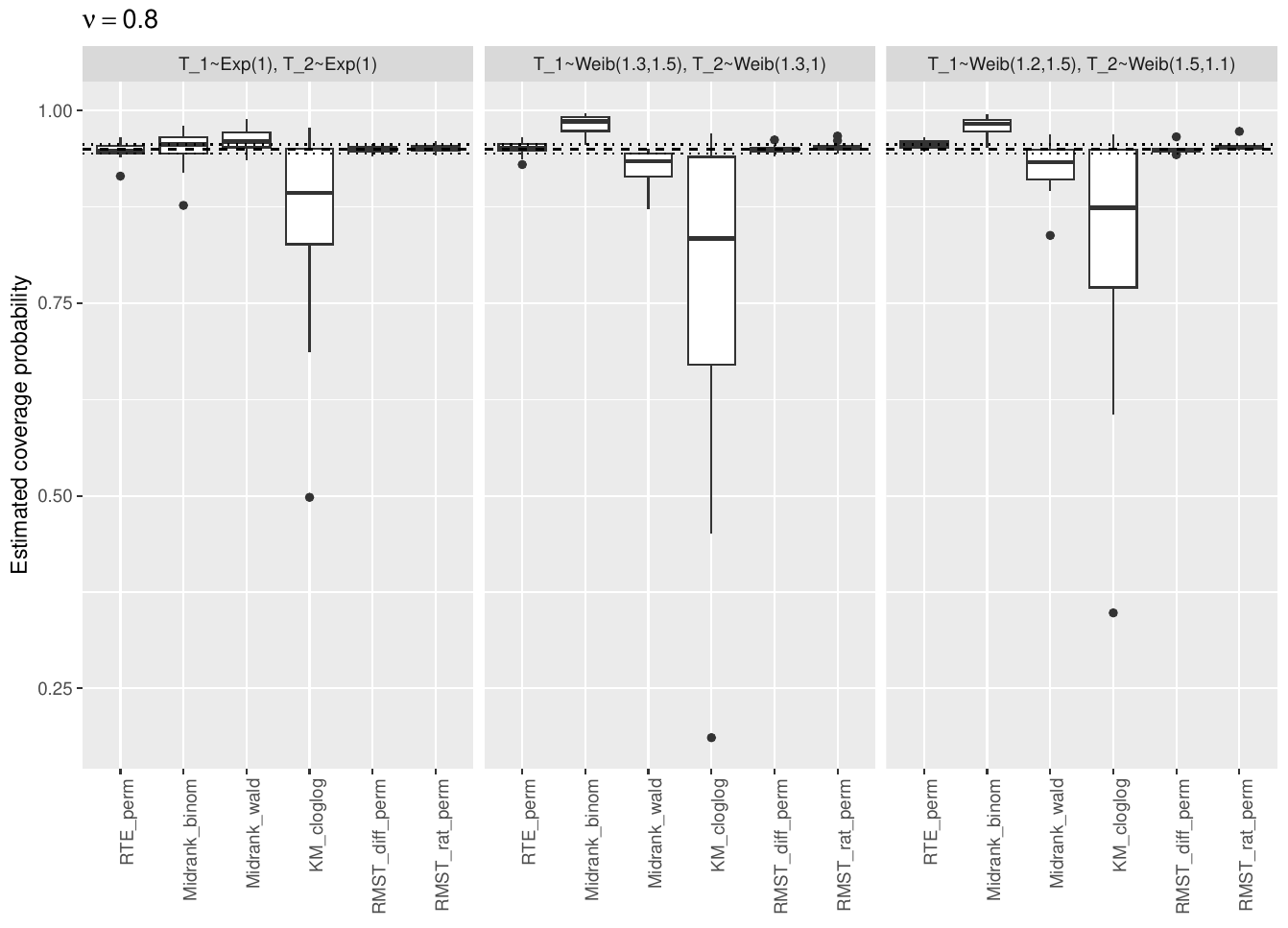}
	\caption{Estimated coverage probabilities for dependence parameter $\nu=0.8$ of left-sided 95\% confidence intervals for $\delta=1$, stratified by marginal distribution. The dashed lines represent the borders of the binomial confidence interval $[94.4\%, 95.6\%]$.}
\label{fig:res_nu0.8_left}
\end{figure}

\begin{figure}[H]
    \centering
	\includegraphics[width=0.75\textwidth]{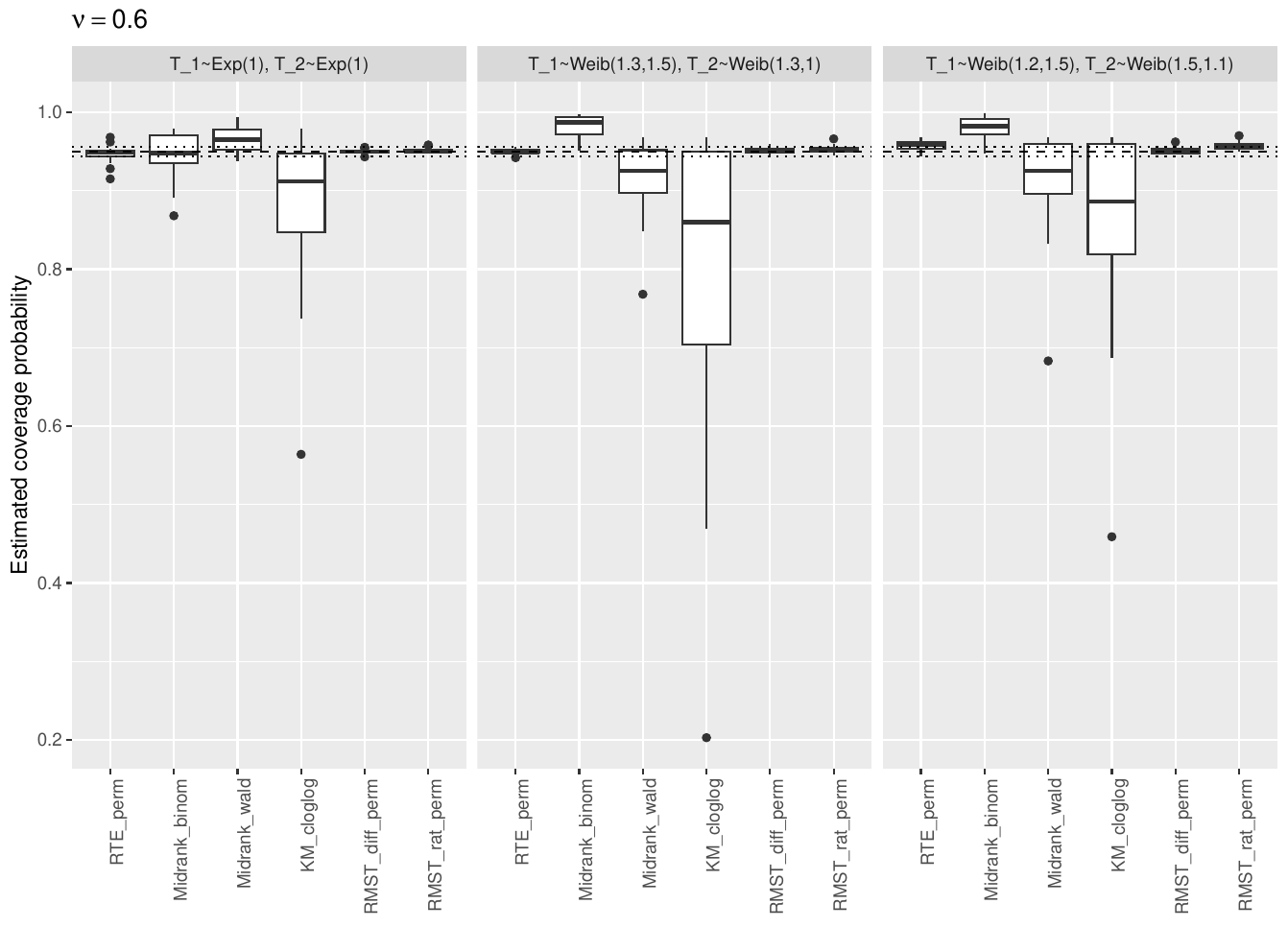}
	\caption{Estimated coverage probabilities for dependence parameter $\nu=0.6$ of left-sided 95\% confidence intervals for $\delta=1$, stratified by marginal distribution. The dashed lines represent the borders of the binomial confidence interval $[94.4\%, 95.6\%]$.}
\label{fig:res_nu0.6_left}
\end{figure}

\begin{figure}[H]
    \centering
	\includegraphics[width=0.75\textwidth]{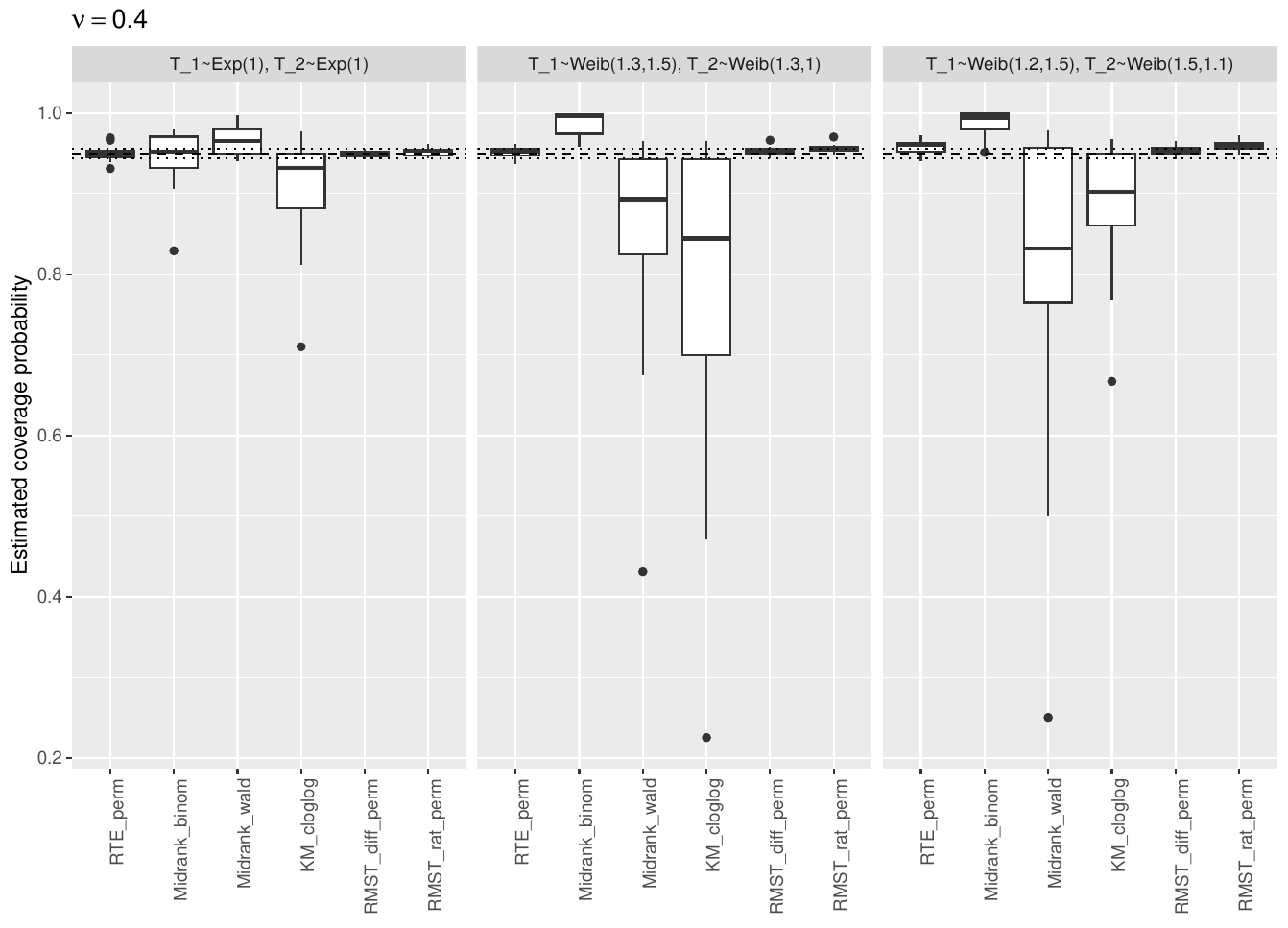}
	\caption{Estimated coverage probabilities for dependence parameter $\nu=0.4$ of left-sided 95\% confidence intervals for $\delta=1$, stratified by marginal distribution. The dashed lines represent the borders of the binomial confidence interval $[94.4\%, 95.6\%]$.}
\label{fig:res_nu0.4_left}
\end{figure}

\subsection{Results by censoring distribution}
\begin{figure}[H]
    \centering
	\includegraphics[width=0.75\textwidth]{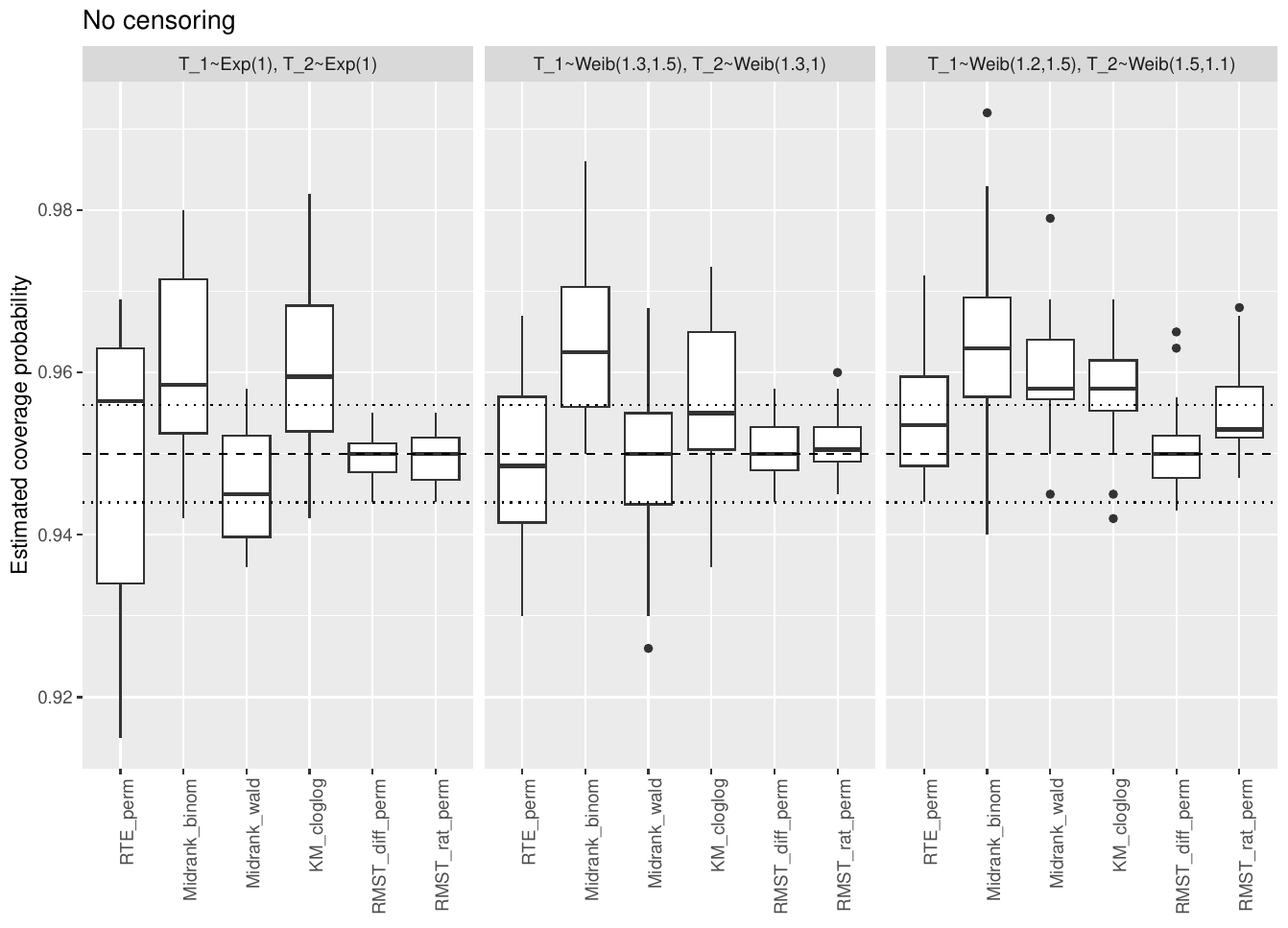}
	\caption{Estimated coverage probabilities for uncensored event times of left-sided 95\% confidence intervals for $\delta=1$, stratified by marginal distribution. The dashed lines represent the borders of the binomial confidence interval $[94.4\%, 95.6\%]$.}
\label{fig:res_nocens_left}
\end{figure}

\begin{figure}[H]
    \centering
	\includegraphics[width=0.75\textwidth]{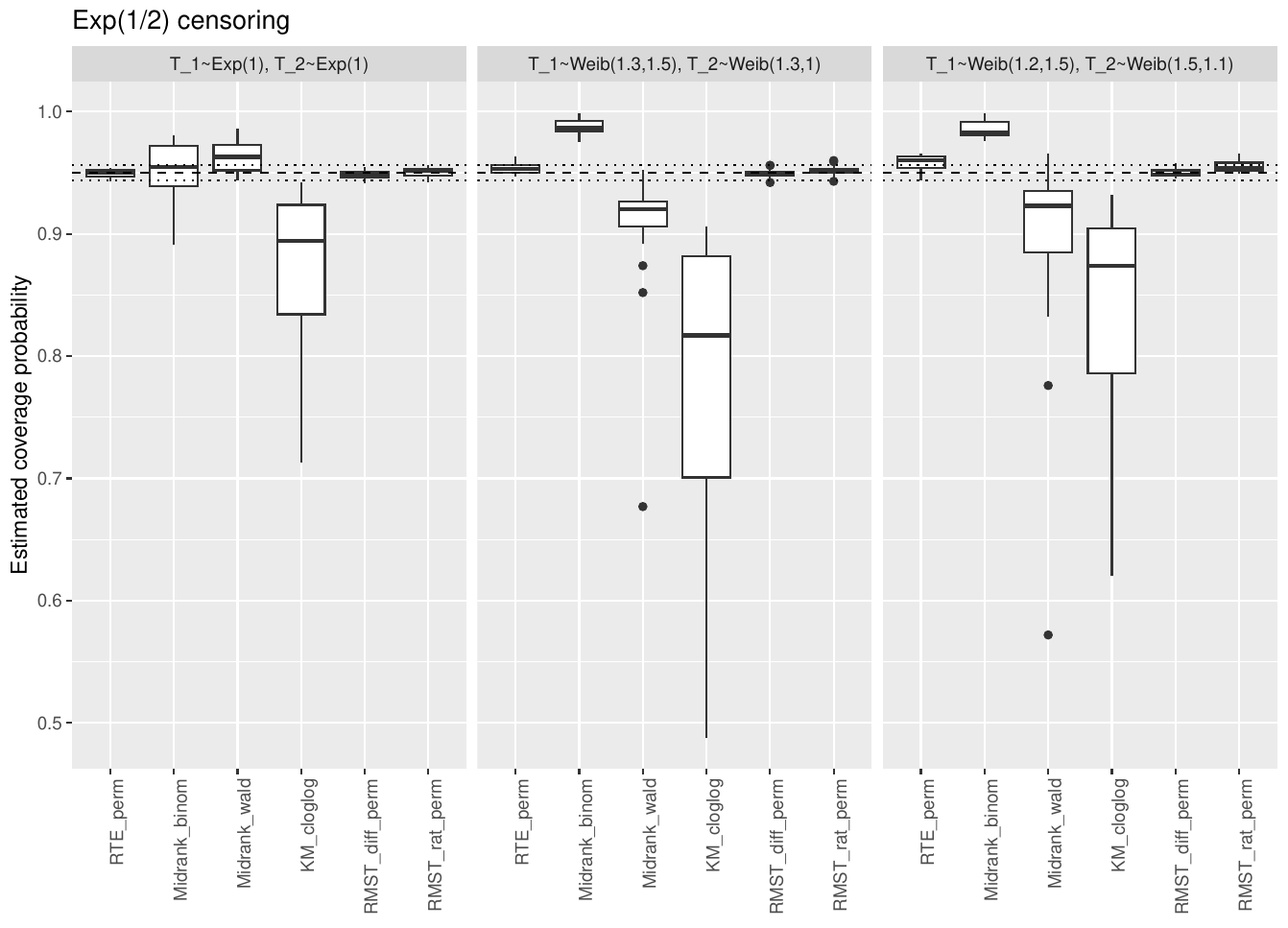}
	\caption{Estimated coverage probabilities for $\text{Exp}(1/2)$ distributed $C_2$ of left-sided 95\% confidence intervals for $\delta=1$, stratified by marginal distribution. The dashed lines represent the borders of the binomial confidence interval $[94.4\%, 95.6\%]$.}
\label{fig:res_exp0.5_left}
\end{figure}

\begin{figure}[H]
    \centering
	\includegraphics[width=0.75\textwidth]{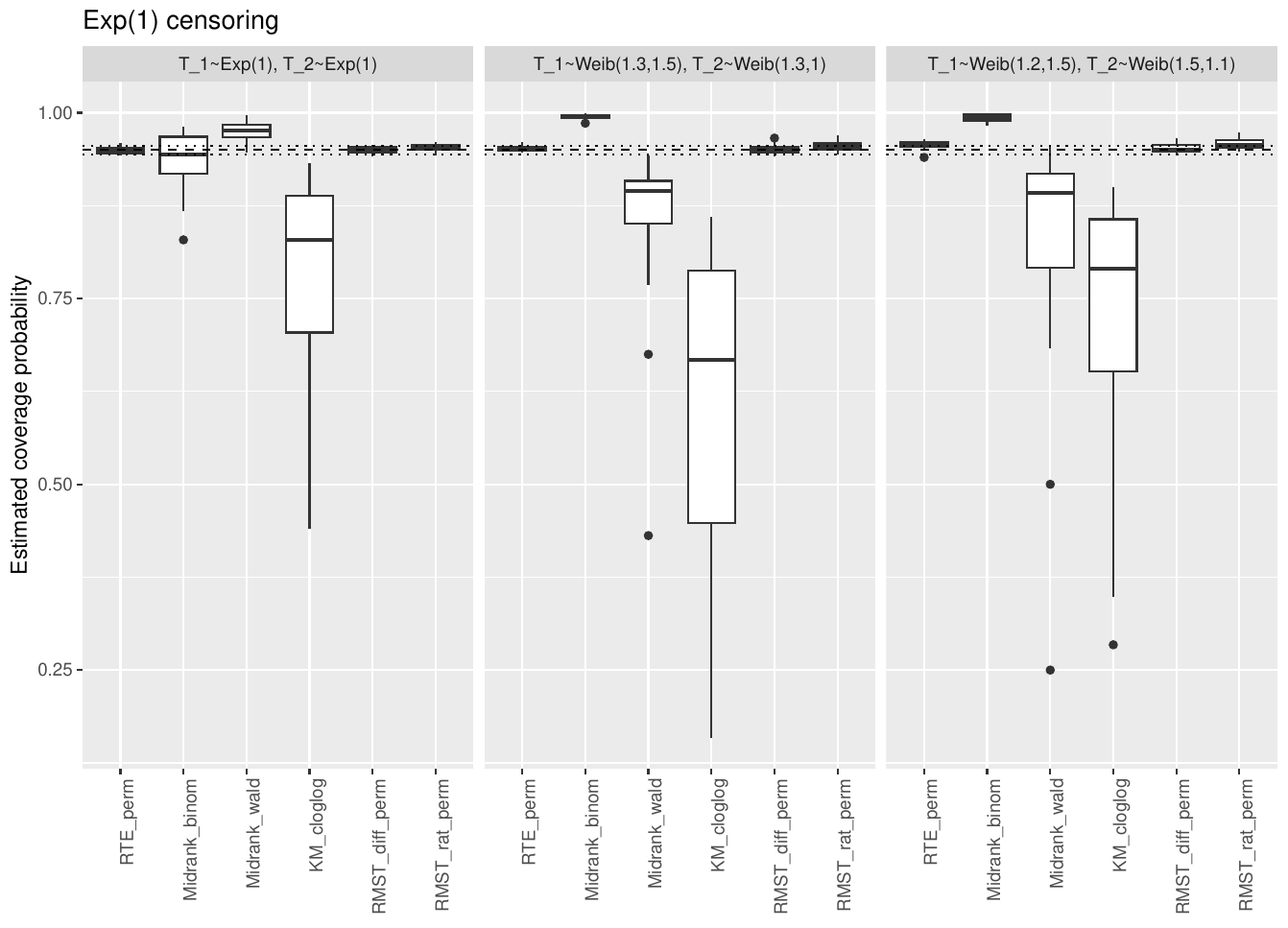}
	\caption{Estimated coverage probabilities for $\text{Exp}(1)$ distributed $C_2$ of left-sided 95\% confidence intervals for $\delta=1$, stratified by marginal distribution. The dashed lines represent the borders of the binomial confidence interval $[94.4\%, 95.6\%]$.}
\label{fig:res_exp1_left}
\end{figure}

\end{document}